# Hierarchical Hidden Markov Model in Detecting Activities of Daily Living in Wearable Videos for Studies of Dementia

Svebor Karaman[(1)], Jenny Benois-Pineau[(1)], Vladislavs Dovgalecs[(2)], Rémi Mégret[(2)], Julien Pinquier[(3)], Régine André-Obrecht[(3)], Yann Gaëstel[(4)] and Jean-François Dartigues[(4)]

*[(1)] LaBRI – University of Bordeaux, 351 Cours de la Libération, 33405 Talence Cedex, France. svebor.karaman@labri.fr, jenny.benois@labri.fr*

*[(2)] IMS – University of Bordeaux, 351 Cours de la Libération, Talence, France. remi.megret@ims-bordeaux.fr, vladislavs.dovgalecs@ims-bordeaux.fr*

*[(3)] IRIT – University of Toulouse, 118 route de Narbonne, 31062 Toulouse Cedex 9, France. pinquier@irit.fr, andre-obrecht@irit.fr*

*[(4)] INSERM U.897 – University Victor Ségalen Bordeaux 2, Bordeaux, France. jean-francois.dartigues@isped.u-bordeaux2.fr, yann.gaestel@isped.u-bordeaux2.fr*

## Abstract

This paper presents a method for indexing activities of daily living in videos acquired from wearable cameras. It addresses the problematic of analyzing the complex multimedia data acquired from wearable devices, which has been recently a growing concern due to the increasing amount of this kind of multimedia data. In the context of dementia diagnosis by doctors, patient activities are recorded in the environment of their home using a lightweight wearable device, to be later visualized by the medical practitioners. The recording mode poses great challenges since the video data consists in a single sequence shot where strong motion and sharp lighting changes often appear. Because of the length of the recordings, tools for an efficient navigation in terms of activities of interest are crucial. Our work introduces a video structuring approach that combines automatic motion based segmentation of the video and activity recognition by a hierarchical two-level Hidden Markov Model. We define a multi-modal description space over visual and audio features, including mid-level features such as motion, location, speech and noise detections. We show their complementarities globally as well as for specific activities. Experiments on real data obtained from the recording of several patients at home show the difficulty of the task and the promising results of the proposed approach.

# 1. Introduction

## a. General problematic for multimedia

With the development and gain of popularity of new smaller, autonomous and capable wearable sensors, we have in the recent years seen a steep increase in the availability of large amounts of new types of data. One can first think on the general public side of User Generated Content [39] captured by smart-devices including camera, microphones, mobile phones and other embedded sensors. These advances have also enabled the development of original applications of multimedia methodologies to new fields, such as health with activity monitoring or memorial tools, sports with performance evaluation using wearable sensors, human-machines interfaces in unconstrained environments using voice or gesture recognition…

Such new data is posing a great challenge to the multimedia community, being of poorer quality and containing less structured content than the majority of digital media documents. Tools and methodologies developed for the analysis of multimedia data have therefore to be adapted and validated specifically in order to be usable in these new applicative contexts. In this paper, we target activity detection using wearable audiovisual capture, and investigate the models, features, and modality fusion aspects required to apply indexing in such challenging real-world conditions.

## b. Motivation

### *i. Societal need for the diagnosis of dementia…*

The application which drives our research is the vast and urgent need for appropriate tools to address the issues raised by the rapid aging of the population of the planet. In particular, elderly people with strong cognitive impairments such as dementia cannot live independently, and the placement in nursing homes becomes unavoidable, with a high cost for the society. One of the major goals in medical research is the early diagnosis of dementia diseases. This would help establish appropriate care giving and postpone placement into specialized institutions, thus reducing the growing costs for the society.

According to results of medical research [1] the first decline in cognitive performances can appear as early as 12 years before the dementia phase and up to 10 years before the individuals become slightly dependent in their Activities of Daily Living (ADLs). To detect the first signs a subject could show, and also to assess the progression of the disease with patients in Mild-Cognitive Impairment (MCI) or established dementia phases, an objective observation of various activities in an everyday life is required for the medical practitioners.

#### *ii. …requiring development of appropriate multimedia indexing tools*

This is why the observation with various types of sensors including video cameras is now entering in clinical practice [29]. The amount of generated video data is usually very large. Indeed, the observation with external cameras in smart homes can last for several hours [2]; the monitoring with wearable sensors can be shorter, but still not browsable by a medical practitioner in the short time allocated to the preparation of an appointment with a patient. Hence, the necessity of automatic recognition of activities of interest is obvious and requires the development of adequate multimedia indexing tools.

In the literature, a large amount of research has been devoted to the recognition of human activities in video recorded with stationary video sensors installed in buildings, e.g. [3]. Some of them are specifically devoted to the elderly people for assessing through observation the degree of autonomy and signs of upcoming dementia [2]. Nevertheless, the "external observation" does not allow for a medical practitioner to observe the actions of the patient in a very detailed way, such as those related to the instrumental activities of daily living. To see how a person uses utensils and food when cooking or how he/she is knitting or washing dishes would require a very dense installation of various sensors at home for each observed patient. Equipping homes with many sensors is a heavy task, which is excessive in the context of a diagnosis, and may not always be well accepted by elderly persons. We are therefore developing an alternative approach, requiring a lighter setup, based on wearable cameras and microphones for activity analysis.

During the last decade various attempts have been made to use a wearable video acquisition set-up for activity monitoring, which we will review in the related work section. As an example, the SenseCam device, which is worn by a person, helps constituting a lifelog that can then be used to rememorize the events for a person with memory impairments [5] [41]. Such wearable video sensors provide an “egocentric” point of view [40], allowing for close-up observation of the activities of a person. They are thus also highly valuable when a fine grained observation of the instrumental activities is required.

### c. Contributions

In this paper we develop the research we first proposed in [4] on activities recognition for the diagnosis and monitoring of dementia. The objective is the indexing of the instrumental activities in audiovisual data recorded with a camera worn by the patient using the paradigm of first-person sensing. We present the methods and algorithms developed for the multi-modal analysis of audio-visual data captured from such a wearable device. Key contributions include:

- a hierarchical two level Hidden Markov Model (HMM) aimed at the detection of activities of daily living, which provides the fusion of multi-modal visual and audio features. These include both low-level descriptors and mid-level features, some of which have been designed for the type of data considered; we show the complementarities of the chosen features globally, and show how specific activities can gain from this;
- a new method for partitioning the data stream into segments, which is designed for continuous video sequences such as those captured by a wearable camera;
- experiments and evaluation on real-life data acquired on both volunteers and patients in the context of the analysis of their activities of daily living.

The paper is organized as follows: in section 2 we review related research and methods. In section 3, we give the motivation for our application of audiovisual lifelogging for the analysis of activities of daily living and describe the architecture of the proposed approach. In section 4, we provide a detailed explanation of the hierarchical HMM. In section 5, we describe our method for partitioning the videos and in section 6 we present our strategy for the extraction of meaningful multimodal features which feed the HMM as observations. Experiments with various configurations of the model and description space are reported in section 7. Finally, we conclude and give perspectives of this work in section 8.

# 2. Related work

## a. Activity recognition in multimedia data

The recognition of events in video has been mostly dedicated to human actions recognition from external cameras [43] with many applications for surveillance and security. This task has been well addressed in the literature, as reviewed in [44, 45]. Tests on standard benchmark databases show almost perfect results [46]. However, the task on these benchmarks is rather simplistic as there is only one low semantic activity in each video which is centered on a single person. Moreover, the audiovisual stream is not fully considered as most of the time only the visual cues are used for the recognition [45].

In more recent works, the video has been considered as a complex multimedia source. Indeed, in addition to the traditional analysis of the visual content, the audio and motion analysis has gained attention in the last years. Thus the problematic of fusion from multiple sources has been addressed [47, 48, 49]. The analysis is often performed on pre-segmented data, i.e. shots in edited videos, where the event of interest has only to be recognized. The application on more generic data imposes methods than can simultaneously segment and recognize from the multimedia stream.

## b. Simultaneous recognition and segmentation on standard multimedia data

The simultaneous segmentation and recognition of videos have been efficiently performed through the Hidden Markov Models formalism. The Hidden Markov Model (HMM) is a statistical model which was first introduced in [8] where its application to speech recognition was presented. The applications of HMMs to video have been first designed for low-level temporal structuring like the method for video segmentation using image, audio and motion content presented in [9], where the HMM states represent the camera motion and the transitions between shots. The richness of application contexts and constraints imposed in various spatio-temporal scenarios yielded a wide range of HMMs. Amongst the variety of HMMs, hierarchical and segmental HMMs turned to be the most popular for modeling activities in video streams.

With regards to the complexity and inherent hierarchical structure in video scenes coming from numerous application domains the classical "flat" HMM is limited. Indeed the structure of video scenes in e.g. sports video can be mapped to more than one HMM. Thus, in tennis videos [10] one HMM can describe a match as a set of "sets", each set can be represented as a set of "games" and each game can be represented as set of "points" etc., up to an elementary events such as "rally", "first missed serve". A vertical link of hierarchy exists between states of the upper level HMM and the lower-level HMMs. This can be represented by the Hierarchical HMM (HHMM) modeling both the hierarchy of events (states) and transitions between them. Usually [14] the hierarchical structure is defined using the bottom-level states as emitting states (where the observation distribution has to be learnt) and high-level states as "internal" states to model the structure of the events. A two-level approach was proposed in [12] where the bottom level is composed of HMMs for feature analysis, and the top level is a stochastic context-free parsing. This model shows improvement in performances of recognition of gestures over a flat HMM. However, one of the main drawbacks of the fully hierarchical models is the higher number of parameters to train (such as the complementary vertical transition probabilities), which induces the need of a large amount of training data.

A major limitation of classical HMMs and derived HHMMs is the invalidity of the assumption of the independence of consecutive observations [14]. In a video this is clearly not the case. Hence the Segmental HMM (SHMM), introduced in [14], addresses the problem of variable length sequences of observation vectors as presented in [15]. The application to video has been, for example, shown for tennis video parsing [16] where thanks to SHMM different modalities can be processed with their native sampling rates and models. Once again, despite the gain in performance, these models have a much higher computational cost and number of parameters than the flat HMM.

Applications to very structured video programs such as news or sports events have been successful. However, the more challenging context of lifelogs analysis arises new problematic and opens a much wider scope of applications for multimedia analysis.

### c. Multimedia analysis of lifelogs

The analysis of image lifelogs captured by the SenseCam device has led to the study of several issues. In [5] and [41], the automatic structuring of scenes in an unsupervised manner is used as an input to present an automatic summary of the daily activities. In [42], the authors extract global information on the various activities and lifestyle of the person by automatically estimating the presence of several semantic concepts in the lifelog thanks to low-level visual features classification.

The WearCam [6] project uses a camera strapped on the head of young children. This setup allows capturing the field of view of the child together with his gaze in order to monitor the impairments in child's development.

The analysis of instrumental activities of daily living using egocentric vision was done in [17]. A HMM model taking into account the objects appearing in specific situated space models related to the person posture was proposed. They established the feasibility of activity recognition based on visual information. Because of the complexity of the task, they did not use real data, but simulated information for the 3D synthetic modeling of the person and the environment. Real data was used in [36], by combining both the wearable camera and inertial sensors as observation for a HMM model for the recognition of standardized actions related to cooking in a test environment. Action recognition was done using only the capture of hands manipulation with a wearable camera in [37]. Here, observation stemming from hand motion templates and external sensors for room transitions are fed to a Dynamic Bayesian Network that infers the activity from a set of predefined sequences of recognized manipulations.

This analysis of the related work shows an increasing interest in the automatic interpretation of multimedia data obtained from wearable sensors. Taking from the successful HMM models for event detection and recognition in generic multimedia data, we will now propose a new method targeted at lifelog data.

## 3. Problem statement and proposed approach

Before developing the proposed algorithmic approach, we provide in this section additional information on the applicative context that motivates it, and that will guide the technical and methodological choices explained later on. After explaining how the audio-visual data is obtained, we will provide more details on the taxonomy of the activities of interest.

### a. Objective

The idea of this research from a medical point of view is to use the video recording in the same way as a clinical test such as MRI or radiography and to get these observations in a stress-less and friendly environment for the patient, while at home. In a target usage scenario, the doctor will ask the paramedical staff to visit the patient with the recording system. Then, the recorded video is automatically processed and indexed by our method off-line. Finally, the doctor will use the video and indexes produced by our analysis to navigate in it and search for the activities of interest. This visual inspection provides clinical elements to be used for the diagnosis of the disease or for the assessment of the evolution of the patient's condition.

The typical recording scenario consists of two stages. At the first stage the patient is asked to walk around his house with the device, which provides a bootstrap video that is representative of the environment. This short video will be annotated in order to build adapted reference models of the places of interest in the patient's home. In the second stage, a longer video is recorded, which is the one to be indexed. During this stage, the patient is asked to perform some of the activities that are part of the clinical evaluation protocols in assessing dementia progress. These activities define the targeted events to be detected by our method.

### b. Data characteristics

Following our preliminary work [4], a vest was adapted to be the support of the camera. The camera is fixed on the shoulder of the patient with hook-and-loops fasteners which allow the camera's position to be adapted to the patient's morphology. This position combined with the wide angle lens of the camera offers a large view field similar to the patient's one. The microphone is integrated in the camera case and records a single audio channel.

Thanks to the low weight and size of the camera and the weight distribution due to the vest, the acceptance of the device is very good. The volunteers have felt no discomfort while wearing it and were able to perform their activities as if the device was not present. An illustration of the device is given in FIGURE 1.

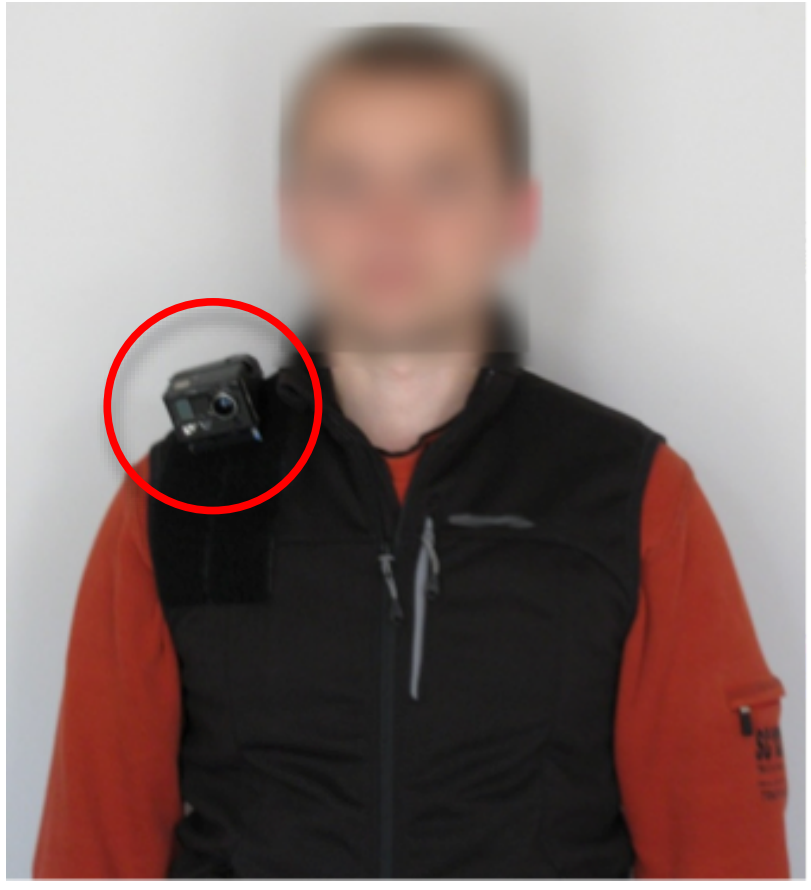

FIGURE 1: The recording device (red circle) fixed on the vest adapted to be the support of the camera.

The videos obtained from wearable cameras are quite different from the standard edited videos in e.g. cinema, commercials, sports or TV programs of other genres. Indeed, edited videos which are usually a target of video indexing methods have smooth motion and are assembled from video shots with discontinuities at the shot borders. In our case, the video is recorded as a long continuous sequence.

Such uninterrupted video is also encountered in surveillance applications, with the important difference that the latter deals with cameras that are stationary or with regular motions, such as PTZ. Videos from wearable cameras suffer from larger and irregular motion. Ego-motion of the patient even of a weak physical magnitude can yield strong changes in the field of view as well as a strong blur. Furthermore, when moving in a natural home environment, the patients face strong light sources, such as windows resulting in saturation of luminance in the field of view. The data variability is also very strong: the same activities are not performed by different patients in the same environment in opposite to “smart homes” setups [2]. Examples of challenging contents from wearable videos are represented in FIGURE 2.

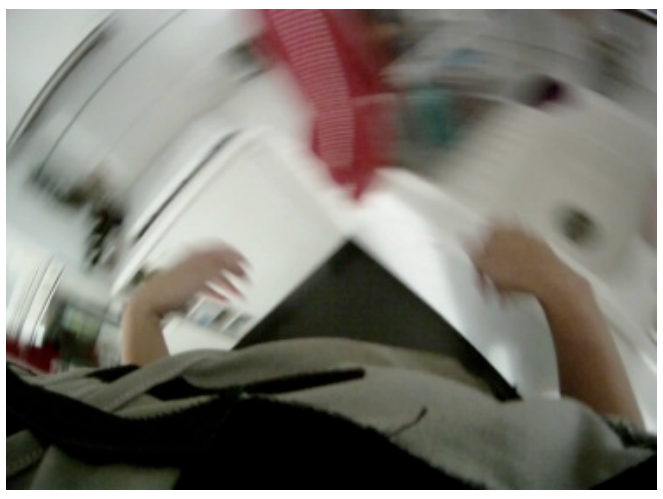

(a) Motion blur due to strong motion.

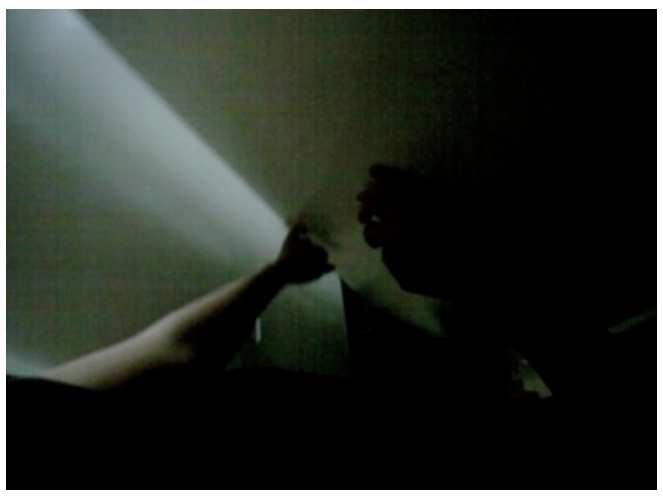

(b) Low lighting while in dark environment.

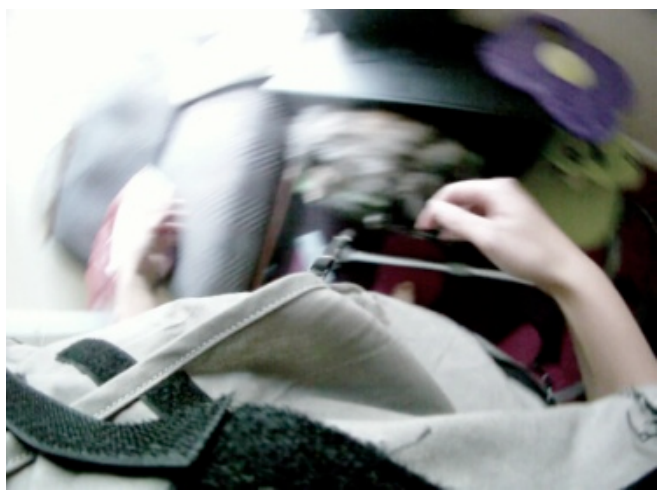

(c) High lighting while facing a window.

FIGURE 2: Examples of frames presenting challenging content for video analysis.

### c. Activities of Daily Living

Up to now, the medical practitioners have been using a paper questionnaire while interviewing the patients and their relatives to determine their ability to correctly perform the following Activities of Daily Living (ADL): “Hoovering”, “Sweeping”, “Washing Clothes”, “Serving”, “Making Coffee”, “Making Snack”, “Hair Brushing”, “Phone”, “TV”, “Knitting”, “Plant Spraying”, “Listening to Radio”, “Wiping Dishes”, “Brushing Teeth”, “Washing Dishes”. These ADL are the target of automatic recognition in order to provide doctors with an efficient navigation tool throughout the video recorded at the patient’s home. In the rest of this paper, we will explain how to detect these activities, and evaluate the recognition performances based on ground truth data. The use of the obtained results inside an interface for browsing the content is the subject of ongoing work with the medical partners, and is out of the scope of this paper.

### d. Architecture of the proposed approach

The proposed approach is based on the multi-modal analysis of the audio-visual content of the data captured with the wearable camera. The problem consists in the recognition of activities that are sequential in time, on the basis of noisy and variable data with some possible constraints on their time scheduling. We resort to HMMs, which proved to be an excellent model for such types of problems [33]. The global architecture is shown in FIGURE 3. The core is the Hierarchical HMM module (HHMM), which takes as input the multi-modal features to detect and segment activities of interest. To account for the specific temporal structure of the wearable video, motion estimation and a new specific temporal partitioning is done on the video data. It is used both as an input for motion description, and as a reference for temporal segmentation of the data stream for the HHMM. Multiple descriptors on video and audio are computed and fused to account for the various types of activities. We will now provide the details of each of these modules.

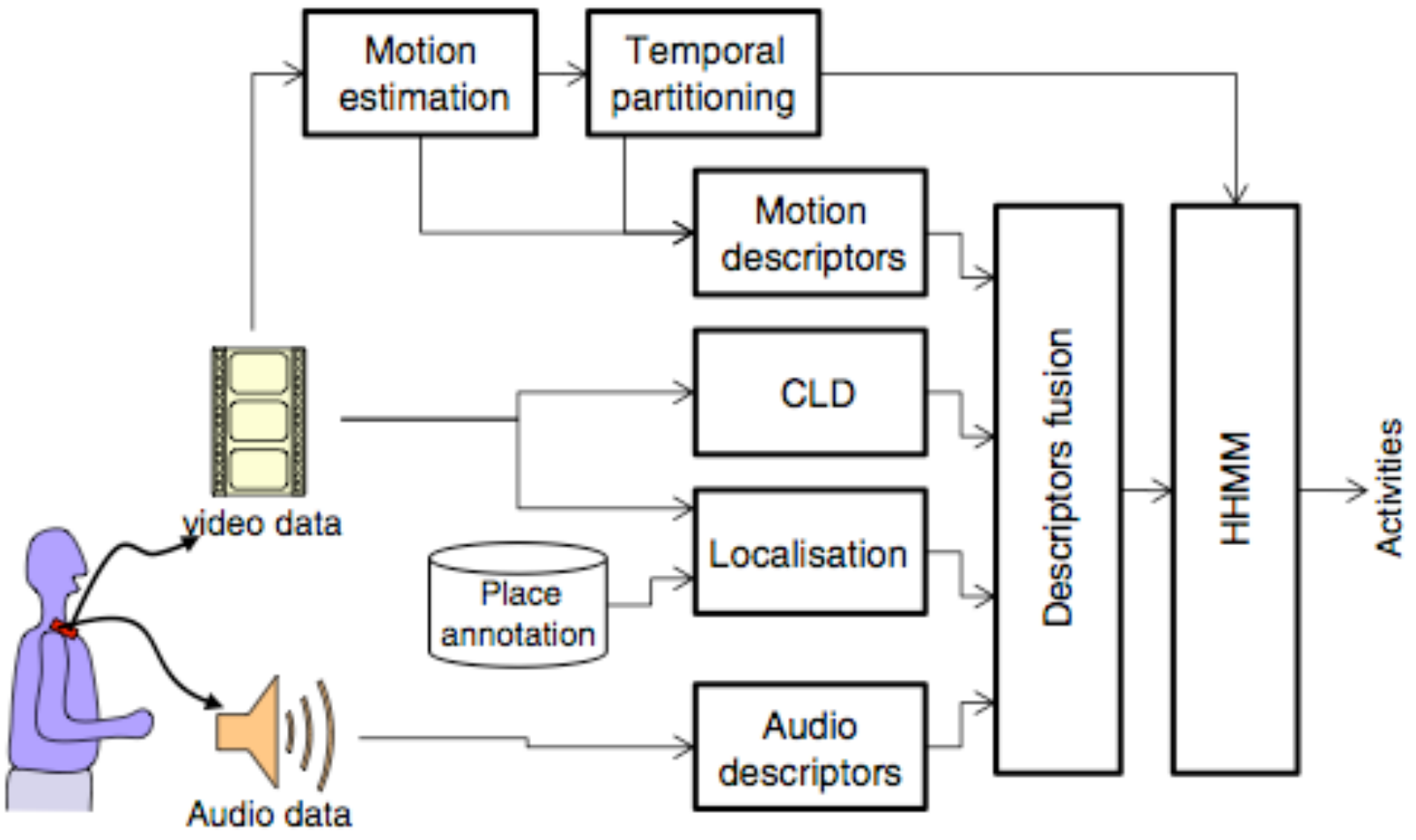

FIGURE 3: Global architecture of the proposed approach.

# 4. Hidden Markov Model for Video Structuring

In this section, we propose an activity recognition model that takes into account both the complexity of our data and the lack of large amount of training data for learning purposes. If we abstract our problem of recognition of daily activities in the video to its simplest core, we can draw an equivalence between an activity and a hidden state of an HMM. This could be achieved by designing a fully connected HMM and training the inherent state-transition probabilities from the labeled data. Unfortunately, the ADL we consider are very heterogeneous and complex, therefore the suggested equivalence between an activity and a hidden state cannot hold.

## a. Two-level Hierarchical HMM (HHMM)

Hence, we propose a two-level Hierarchical HMM (HHMM). The activities that are meaningful to the medical practitioners are encoded in the top-level HMM, the set of possible states is thus defined accordingly. We also introduce a reject state "None" to model non-meaningful observations from the doctors' point of view. Defined as such, the top-level HMM contains the transitions between "semantic" activities including the reject class. A bottom-level HHM models an activity with $m$ non-semantic states, as in [17]. The number of non-semantic states associated to a semantic state is fixed to 3, 5 or 7 for ADL states and to 1, 3, 5 or 9 for the reject class "None" in our experiments. A simplified illustration of the overall structure of the HHMM is presented in FIGURE 4, for the case of 3 states at the bottom level.

### b. Top-level HMM

The top-level HMM represents the relations between the actions of interest, which are the ADL defined by the medical practitioners and another activity for all the irrelevant actions named "None". We denote the set of states at this level by $Q^1 = \{q_1^1, \dots, q_{n_1}^1\}$ and the transition matrix by $A^1 = (a_{ij}^1)$, where $n_1$ is the number of activities. In this work, no constraints were specified over the transitions between activities as such restrictions are very difficult to know *a priori* when addressing a larger set of activities and when analyzing a large set of videos where the physical constraints of each patient's house are different. Moreover, the ADL a patient is asked to fulfill depend very much on his condition and their sequencing cannot be fixed for all patients in the same way. Hence, we design the top-level HMM as a fully connected one. We consider equi-probable transitions from activities states to one another, hence $\forall i,j: a_{ij}^1 = \frac{1}{n_1}$. These transitions are fixed *a priori* and are not re-estimated during the learning phase. The states of the top-level HMM modeling activities are denoted in FIGURE 4 by $Act_i$. The probabilities of vertical transitions are denoted by $a_{start_i,j}^2$ and are modeled by using a virtual node $Start_i$; they are the equivalent to the transitions denoted by $\pi^{Act_i}(q_j^2)$ in the HHMM formalism of [13].

### c. Bottom-level HMM

Most of the activities, defined in the above section, are complex and could not easily be modeled by one state. For each activity $q_i^1$ in the top-level HMM a bottom-level HMM is defined with the set of states $Q_i^2 = \left\{q_{i1}^2, \dots, q_{n_i^2}^2\right\}$ with $n_i^2 = 3, 5 \text{ or } 7$ for ADL states and $n_i^2 = 3, 5, 7 \text{ or } 9$ states for the reject class "none" in our experiments. The state transition matrices $A_i^2$, for $i = 1, \dots, n_1$ also correspond to a fully connected HMM: $a_{ikl}^2 \neq 0$, at initialization, for $k = 1, \dots, n_i^2$ and $l = 1, \ldots, n_i^2$. For the video stream not to be over-segmented the loop probabilities $a_{ikk}^2$ have to be initialized with stronger values than other transition probabilities: $a_{ikk}^2 > a_{ikl}^2, \forall\, k \neq l$. Activities are more likely to involve several successive observations rather than just one: this explains the choice for such a higher loop probability.

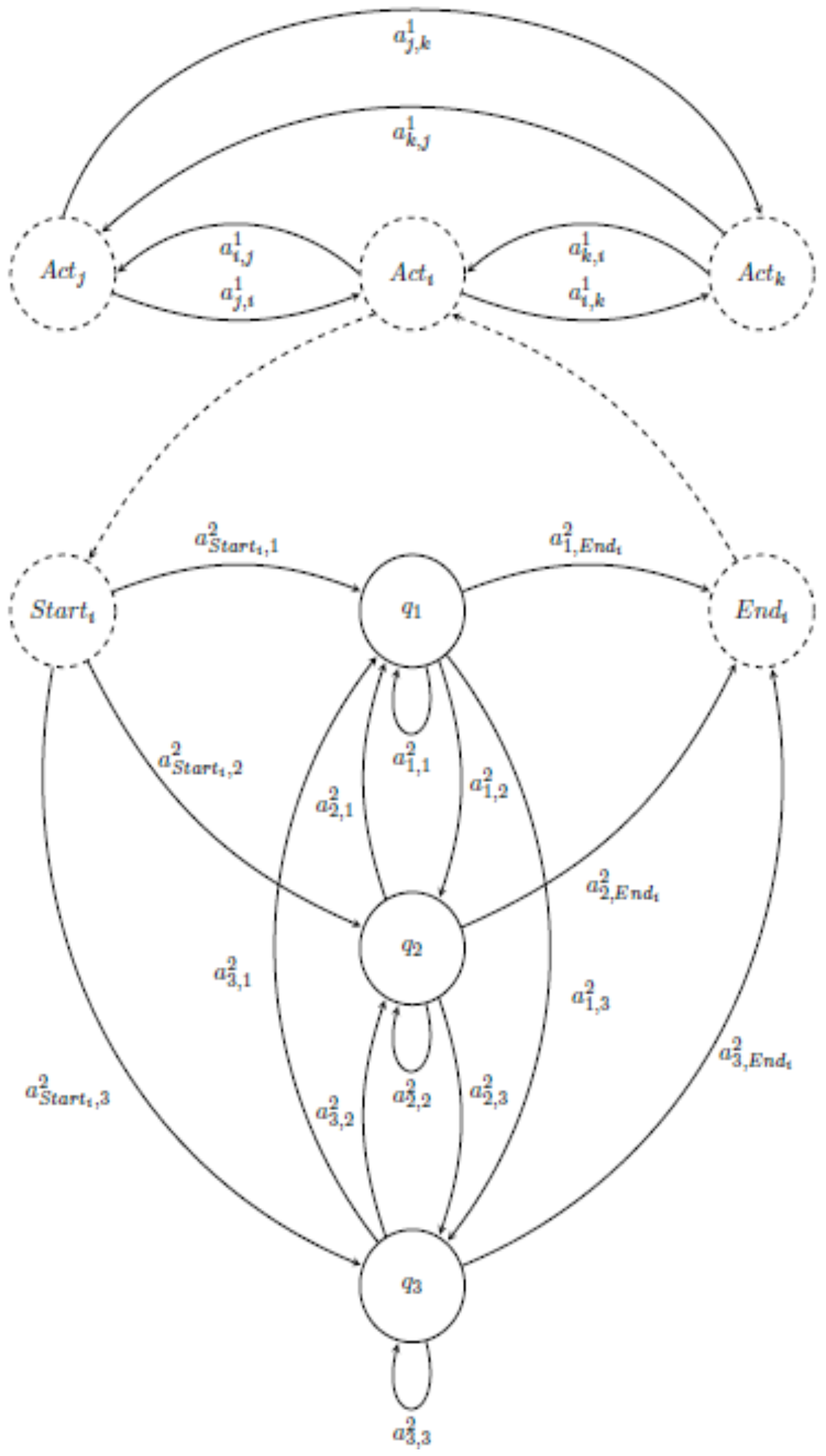

FIGURE 4: Generic structure of our two-level HHMM for modeling activities of a patient. Act. Activities, q: emitting states. Dashed circled states are non emitting states. In the experiments, this structure is defined for 16 or 23 ADL with 3, 5 or 7 non-semantic states.

At the bottom level, we consider a continuous HMM that models observations probability with a Gaussian Mixture Model (GMM), i.e. each non semantic state models the observation vector $o$ by a GMM. In our model, we consider a diagonal covariance matrix. The number of states at the bottom level is fixed and will not be changed during the learning process. The GMM and the transitions matrix of all the bottom-level HMMs are learned using the classical Baum Welsh algorithm [8] with labeled data corresponding to each activity. In our implementation of the designed two-level HHMM, we used the HTK library [30], which is available as open-source software [7].

For the recognition, the Viterbi algorithm is used. The HTK implementation makes efficient use of the "token passing" paradigm to implement a beam pruned Viterbi search [31].

# 5. Partitioning into analysis units

The video structuring will rely on an analysis unit. We want to establish a minimal analysis unit which is more relevant than a single video frame. The objective is to segment the video into the different "viewpoints" that the patient provides by moving throughout his home. In contrast to the work in [5] where the description space is based on a fixed key-framing of the video, our goal is to use the motion of the patient as one of the features. This choice corresponds to the need to distinguish between activities of a patient which are naturally static (e.g. reading) and dynamic (e.g. hoovering). This segmentation into "viewpoints" of our long uninterrupted video sequences may be considered as an equivalent to partitioning of edited video sequences into shots. We now detail the designed motion-based segmentation of the video.

## a. Temporal integration of global motion

Since the camera is worn by the person, the global motion observed in an image plane can be called the ego-motion. We model the ego-motion by the first order complete affine model and estimate it with a robust weighted least squares by the method we reported in [18]. The parameters of (1) are computed from the motion vectors extracted from the compressed video stream (H.264 in the current recording device) where one motion vector $\vec{d_i} = (dx_i, dy_i)$ is extracted for each image block and is assumed to follow the global motion model

$$\begin{pmatrix} dx_i \\ dy_i \end{pmatrix} = \begin{pmatrix} a_1 \\ a_4 \end{pmatrix} + \begin{pmatrix} a_2 & a_3 \\ a_5 & a_6 \end{pmatrix} \begin{pmatrix} x_i \\ y_i \end{pmatrix} \quad (1)$$

with $(x_i, y_i)$ being the coordinates of the block center.

To split the video stream into segments, we compute the trajectories of the frame corners using the global motion estimation previously presented. For each frame, the distance between the initial and the current position of a corner is calculated. We denote by $w$ the image width and by $s$ a threshold on the frame overlap rate. A corner is considered as having reached an outbound position once it has had a distance greater than $s * w$ from its initial position in the current segment. These boundaries are represented by green and red (when the corner has reached an outbound position) circles in FIGURE 5.

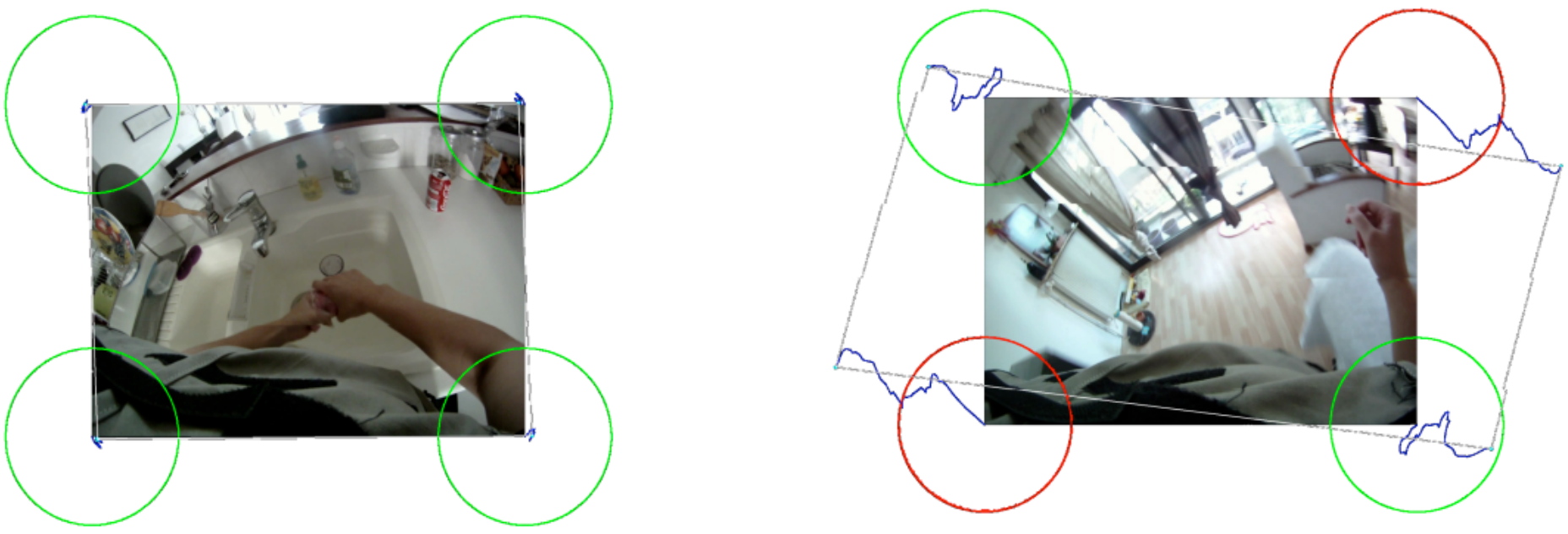

(a) Corner trajectories while the person is static. (b) Corner trajectories while the person is moving.

FIGURE 5: Example of corners trajectories.

### b. Definition of Segments

Each segment $S_k$ corresponds to a temporal interval $S_k = [t_k^{min}, t_k^{max}]$ which aims at representing a single “viewpoint”. The notion of viewpoint is clearly linked to the threshold $s$. The latter defines the minimal proportion of the first frame of a segment, which should be contained in all its frames. This threshold was fixed to 0.2 according to performance results of previous experiments [32]. We define the following rules: a segment should contain a minimum of 5 frames and a maximum of 1000 frames. These boundaries on segment duration are defined to avoid an over-segmentation of the video by setting a minimal duration corresponding to a sixth of second, and to avoid having a long static activity represented by a single segment [32]. The end of the segment is the frame corresponding to the time when at least 3 corners have reached at least once an outbound position. The key frame is then chosen as the temporal center of the segment, as illustrated in FIGURE 6. Hence the estimated motion model serves two goals: i) the estimated motion parameters are used for the computation of dynamic features in the global description space, and ii) the key frames extracted from motion-segmented "viewpoints" are the basis for extraction of spatial features.

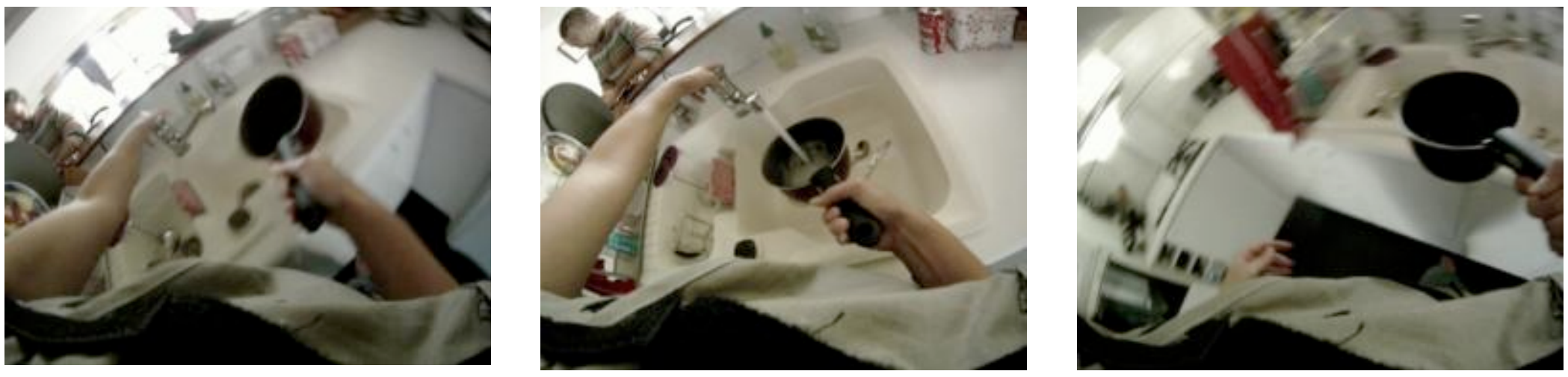

FIGURE 6: An example of key frame (center) with the beginning (left) and ending (right) frames of the segment.

# 6. Multi-modal features

Given the list of activities of interest, no single feature can resume all relevant information from the raw data. We therefore introduce the fusion of several features, combining visual descriptors (motion and static visual features), and audio descriptors. Each feature is designed to bring specific and complementary information about the observed activity.
We first introduce descriptors that characterize the motion within the video recorded, then define the audio analysis and finally present static descriptors that gather the context of the patient's environment. Finally, the fusion of all these features will be presented.

## a. Motion description

Motion contains relevant information to characterize an activity. Indeed, the camera being worn by the patient, the global motion corresponds to the ego-motion. Thus, the parameters of the global motion model are directly linked to the instantaneous displacement of the patient and can help to distinguish between static or dynamic activities. Since instantaneous motion may be limited to describe highly dynamic activities such as the hoovering, we will also propose a description of motion history that characterizes the dynamics on a longer term. Finally, the local residual motion is also important and may characterize a moving object or an interaction with an object. Therefore, a set of descriptors capturing several complementary properties of the observed motion will be defined.

### i. *Global and Instant Motion*

The ego-motion is estimated by the global motion analysis presented in section 5.a. The parameters $a_1$ and $a_4$ are the translation parameters. We limit our analysis to these parameters, since as in the case of wearable cameras, they better express the dynamics of the behavior, and pure affine deformation without any translation is practically never observed.

The instant motion histogram is defined as the histogram of the log-energy of each translation parameter $H_{tpe}$, as expressed in (2), defining a step $s_h$ and using a log scale. Since this histogram characterizes the instant motion it is computed for each frame. This feature is designed to distinguish between "static" activities e.g. "knitting" and dynamic activities, such as "sweeping".

$$H_{tpe,j}[i] = \begin{cases} 1 \text{ iff } a_j(t) \in B_i \\ 0 \text{ otherwise} \end{cases} \text{ with the bins } B_i \text{ defined as}$$

$$\begin{array}{llll} a_j \in B_1 & \text{iff} & \log(a_j{}^2) < i * s_h & \text{for } i = 1 \\ a_j \in B_i & \text{iff} & (i-1) * s_h \le \log(a_j{}^2) < i * s_h & \text{for } i = 2..N_e - 1 \\ a_j \in B_{N_e} & \text{iff} & \log(a_j{}^2) \ge (i-1) * s_h & \text{for } i = N_e \end{array}$$

Eq. (2): Translation parameter histogram, associated to a segment $S_k$, where $a_j$ is either $a_1$ or $a_4$.

The feature for a video segment $S_k$ is an averaged histogram on all its frames: $\overline{H_{tpe,J}}$, j=1,4 for horizontal and vertical translations parameters, respectively $a_1$ and $a_4$. The global instant motion feature is the concatenation of both: $\overline{H_{tpe}} = \left(\overline{H_{tpe,1}}, \overline{H_{tpe,4}}\right)$.

We denote by $H_{tpe}(x) = H_{tpe,1}$ the histogram of the log-energy of horizontal translation, and by $H_{tpe}(y) = H_{tpe,4}$ the histogram of the log energy of vertical translation observed in image plane. The number of bins $N_e = 5$ is chosen empirically. This configuration is fixed for all the experiments and has been determined on a corpus of six videos. The threshold $s_h$ is chosen in such a way that the last bin corresponds to the translation of the image width or height respectively.

### ii. History of Global Motion

Another element to distinguish static and dynamic activities is the "motion history". On the contrary to the instant motion, we design it to characterize long-term dynamic activities, such as walking ahead, vacuum cleaning, etc. The estimation of this is done by computing a "cut histogram" $H_c$. The i-th bin of this histogram contains the number $H_c(i)$ of cuts (according to the motion based segmentation presented in section 5.a) that happened in the last $2^i$ frames, see FIGURE 7. The number of bins $N_c$ is defined as 8 in our experiments providing a history horizon of 256 frames. This represents almost 9 seconds of our 30 fps videos. The history horizon was chosen to be the highest power of two lower than the minimal average duration of an activity. Such a definition is a good trade-off between long term history and potential overlapping of activities. Thus defined, the cut histogram is associated to each frame in the video. The descriptor associated to a full segment is the average of the cut histograms of the frames belonging to the segment.

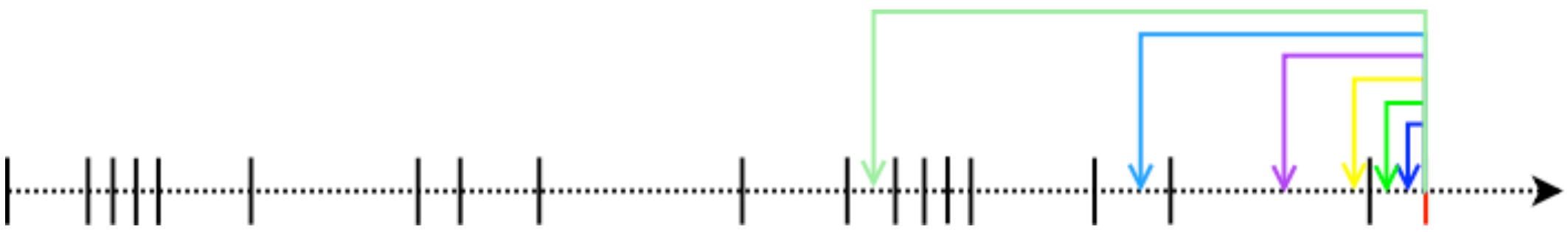

FIGURE 7: The number of cuts (black lines) is summed to define the value of each bin. In this example: $H_c[1]=0$, $H_c[2]=0$, $H_c[3]=1$, $H_c[4]=1$, $H_c[5]=2$, $H_c[6]=7$.

### iii. Local Residual Motion

All the previous motion descriptors focus on the global motion which is very important as it provides a characterization of the ego-motion. However, the residual motion may reveal additional information, such as the occurrence of a manual activity or the presence of a moving object or a person in the visual field of view of the patient. We introduce a descriptor which is computed on each block of a 4×4 grid partitioning of an image. The value of the local residual motion $\boldsymbol{RM_b}$ representing a block $\boldsymbol{b}$ is defined (see equation (3)) as the Root Mean Square (RMS) of the difference $\overline{\Delta d}_{k,l} = \left(\Delta dx_{k,l}, \Delta dy_{k,l}\right)^T$ between motion vector extracted from compressed stream and the one obtained from the estimated model (1). The residual motion descriptor RM obtained by concatenating the values for the 4×4 blocks covering the whole image has therefore 16 dimensions.

$$RM_b = \sqrt{\frac{\sum_{k=1,l=1}^{k=N,l=M}(\Delta \boldsymbol{dx}_{k,l}{}^2 + \Delta \boldsymbol{dy}_{k,l}{}^2)}{N * M}} \qquad (3)$$

Eq. (3): Residual Motion value for block $b$ of width $N$ and height $M$.

## b. Audio

The particularity of our contribution in the design of a description space consists in the use of low-level audio descriptors. Indeed, in the home environment, there are a lot of significant sounds: ambient TV audio track, noise produced by different objects that the patient is manipulating, conversations with the persons, etc. All these sounds are good indicators of an activity and its location. In order to characterize the audio environment, different sets of features are extracted. Each set is characteristic to detect a particular sound: speech, music, noise and silence with a constant objective: having robust audio features without training (or knowledge) of the audio context [26]. The whole system is illustrated on FIGURE 8.

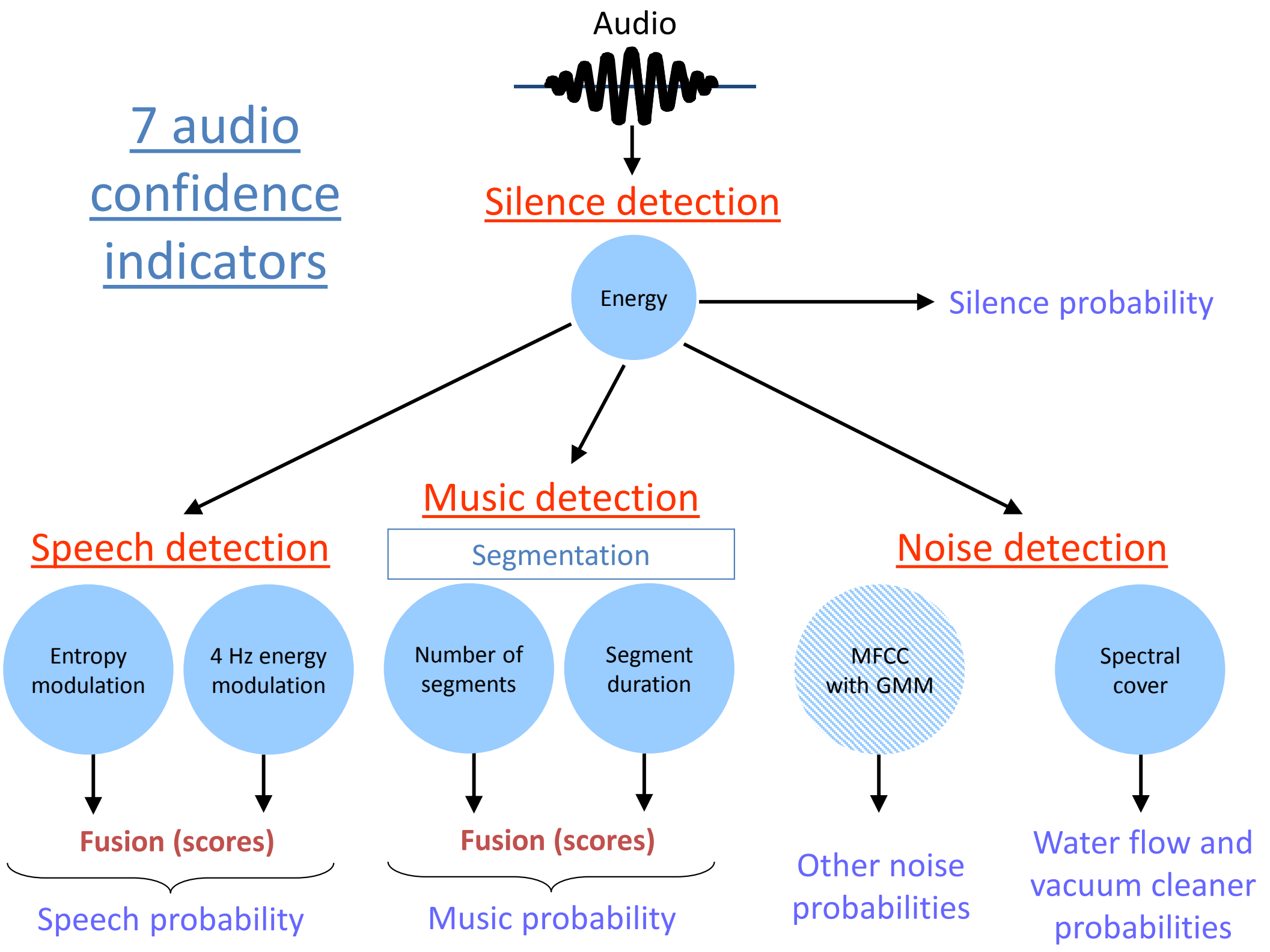

FIGURE 8: Extraction system for audio confidence indicators (probabilities).

Energy is used for silence detection. 4 Hertz energy modulation and entropy modulation give voicing information, being specific to the presence of speech. The number of segments per second and the segment duration, resulting from a "Forward-Backward" divergence algorithm [25], are used to find harmonic sound, like music. To precise the noise component, we propose an original low level descriptor called "spectral cover" that allows recognizing two specific sound events: water flow and vacuum cleaner [34]. These features are based on a threshold to determine if the sound is present or absent. Then, for each sound type, the confidence measure (probability) is the proportion of detected events by segment (defined in section 5.b).

This audio system is complemented with a classical approach (MFCC-based GMM) to have more noise information: percussion and periodic sounds (examples: footstep, home appliance, applause, laugh, etc.). These two features provide, like the previous ones, probabilities (see also FIGURE 8). Model parameters were not learned on the IMMED corpus but on a radio corpus from the ESTER2 evaluation campaign for the Sound Event Segmentation (SES) task [35]. Finally, the complete set of audio descriptors is composed of 7 possible events: speech, music, noise, silence, periodic sounds, water flow and vacuum cleaner.

### c. Static descriptors

Static descriptors aim to characterize the instantaneous state of the patient within his/her environment. The first static descriptor is the localization estimation. As many activities are linked to a specific location e.g. cooking in the kitchen, this feature is essential to provide a context for the activities. The second defines the local spatial and color environment using the MPEG-7 descriptor "Color Layout". This descriptor aims at capturing the spatial and color organization of local pattern when the patient is in a more specific situation, such as facing a sink or a gas cooker for example.

#### *i. Localization*

We use the Bag of Visual Words method [19] for representing an image as a histogram of visual words. Low level visual information contained within an image is captured using local features SURF [22] descriptors. Descriptors are quantized into visual words using a pre-built vocabulary which is constructed in a hierarchical manner [20]. The Bag of Words vector is built by counting the occurrence of each visual word. Due to rich visual content, the dimensionality of such histograms is very high (we used a 1111 word dictionary in our context). A kernel based approach based on the SVM classifier [24] was therefore chosen to obtain location estimates. The histograms were compared with the intersection kernel, which is adapted to such features. In practice, the feature extraction step can be done without annotation, and can be run as a preprocessing routine. Dimensionality reduction through non-linear Kernel PCA [21] with intersection kernel was included in this routine to reduce the size of the stored descriptors to several hundred linear dimensions [23]. Classification was then applied directly on these simplified descriptors. A one-vs-all approach was used to address the multi-class classification problem. The final location was represented as a vector, containing a 1 for the detected class (within "bathroom", "bedroom", "kitchen", "living room", "outside", "other" or "reject"), and 0 for the other classes.

#### *ii. Spatial and color description*

Using the extracted key frames representing each segment, a simple description of the local spatial and color environment is expected. In this choice we seek for the global descriptors which characterize the color of frames while still preserving some spatial information. The MPEG-7 Color Layout Descriptor (CLD) proved to be a good compromise for both [27]. It is a vector of DCT coefficients computed on a roughly low-passed filtered and sub-sampled image. We compute it on each key frame and retain 6 parameters for the luminance and 3 for each chrominance as in [28]. This descriptor provides a coarse yet discriminative visual summary of the local environment.

### d. Descriptors fusion

For the description of the content recorded with the wearable camera we designed three description subspaces: the "dynamic" subspace has 34 dimensions, and contains the descriptors $(H_{tpe}(x), H_{tpe}(y), H_c, RM)$; the "audio" subspace contains the $k = 7$ audio descriptors $p = (p_1, \ldots, p_k)$; the "static" subspace contains 19 coefficients, more precisely $l = 12$ CLD coefficients $C = (c_1, \ldots, c_l)$ and m = 7 localization coefficients $L = (l_1, \ldots, l_m)$. A reminder of the descriptors definitions is given in Table 1.

We design the global description space in an "early fusion" manner concatenating all descriptors in an observation vector $o$ in $\mathbb{R}^n$ space with n = 60 dimensions when all descriptors are used, thus, the designed description space is inhomogeneous. We will study the completeness and redundancy of this space in a pure experimental way with regard to the indexing of activities in Section 7, by building all the possible partial fusions.

**Table 1**: Descriptors definitions.

| | | |
|---|---|---|
| Dynamic | $H_{tpe}$ | Instant global motion descriptor |
| | $H_c$ | History of global motion descriptor |
| | $RM$ | Local motion descriptor |
| Static | CLD | Color Layout Descriptor (MPEG-7) |
| | $Loc$ | Localization estimation within the 7 localization classes |
| Audio | $Audio$ | Concatenation of the 7 audio probabilistic features |

# 7. Experiments

## a. Corpus

The experiments are conducted on a corpus of videos recorded with our wearable device by patients in their own houses. A video recording is of an average duration of 40 minutes and contains approximately 10 activities; not all activities are present in each video. Each video represents an amount of 50000 to 70000 frames, which induces hundreds to a thousand segments according to our motion-based temporal segmentation, see section 5. The description spaces are built using each descriptor separately and with all possible combinations of descriptors where order is not considered. Therefore, a total of 63 different descriptions spaces are considered.

The experiments are conducted in two stages. First, on a corpus of 5 videos recorded with 5 different patients. The aim of this first experiment is to analyze the overall performances of all the descriptors combinations and of the HMM configurations. The influence of the proposed motion-based temporal segmentation is also discussed in the first experiment. The second experiment uses a subset of all the descriptors combinations, selected as the 13 best performances. The HMM configurations are also limited to those who have shown the best performances on the first experiment. In this experiment, the corpus is larger as it contains 26 videos. The latter constitutes a unique corpus which has been recorded on healthy volunteers and patients during two years since the beginning of the research. The performance analysis is also two-fold: we evaluate it in terms of global accuracy and for singular activities.

### b. Evaluation metrics

To evaluate the *overall performance* of the proposed model we used the global accuracy metric, which is a ratio between the number of correct estimations and the total number of observations. Any misclassification of an activity, which will correspond to a *false negative* (FN) with regard to the ground truth activity and to a *false positive* (FP) with regard to the detected activity, will decrease the global accuracy metric.

**Table 2**: Evaluation metrics.

| | |
|---|---|
| $precision = \frac{TP}{TP + FP}$ | $recall = \frac{TP}{TP + FN}$ |
| $accuracy = \frac{TP + TN}{TP + FP + TN + FN}$ | $F - score = \frac{1}{{}^{1}/_{precision} + {}^{1}/_{recall}}$ |

The precision, recall and F-score metrics are used for the evaluating the recognition performances of a particular activity. *True positives* (TP), *true negatives* (TN), *false positives* (FP) and *false negatives* (FN) values correspond to the correct detection, correct absence, misdetection and missed detection respectively for a given activity. The definitions of the used metrics are presented in Table 2.

### c. Learning and testing protocol

The experiments are conducted in a leave-one-out cross validation scheme, i.e. the HMMs are learned using all videos except one which is used for testing. The results are presented in terms of global accuracy of recognition averaged over the cross validation process.

The training is performed over a sub sampling of smoothed data extracted from the video frames. The smoothing substitutes the value of each frame descriptor by the average value on the 10 surrounding frames, then one of ten samples is selected to build ten times more learning sequences. The testing has been done on frames or segments of the last video. The label of a segment is derived from the ground truth as being the activity having the more frames labeled within the segment boundaries.

In the first experiment presented here, the bottom level HMM of each activity has 3 or 5 states. For a given evaluation, all activities have the same number of states, except the “None” which may be modeled with more or fewer states, here 9 or only one. We denote by “Xstates” the configuration where all activities have the same number (X) of states and “XstatesNoneYStates” the configuration where genuine activities have X states and the reject class “None” has Y states. Inside

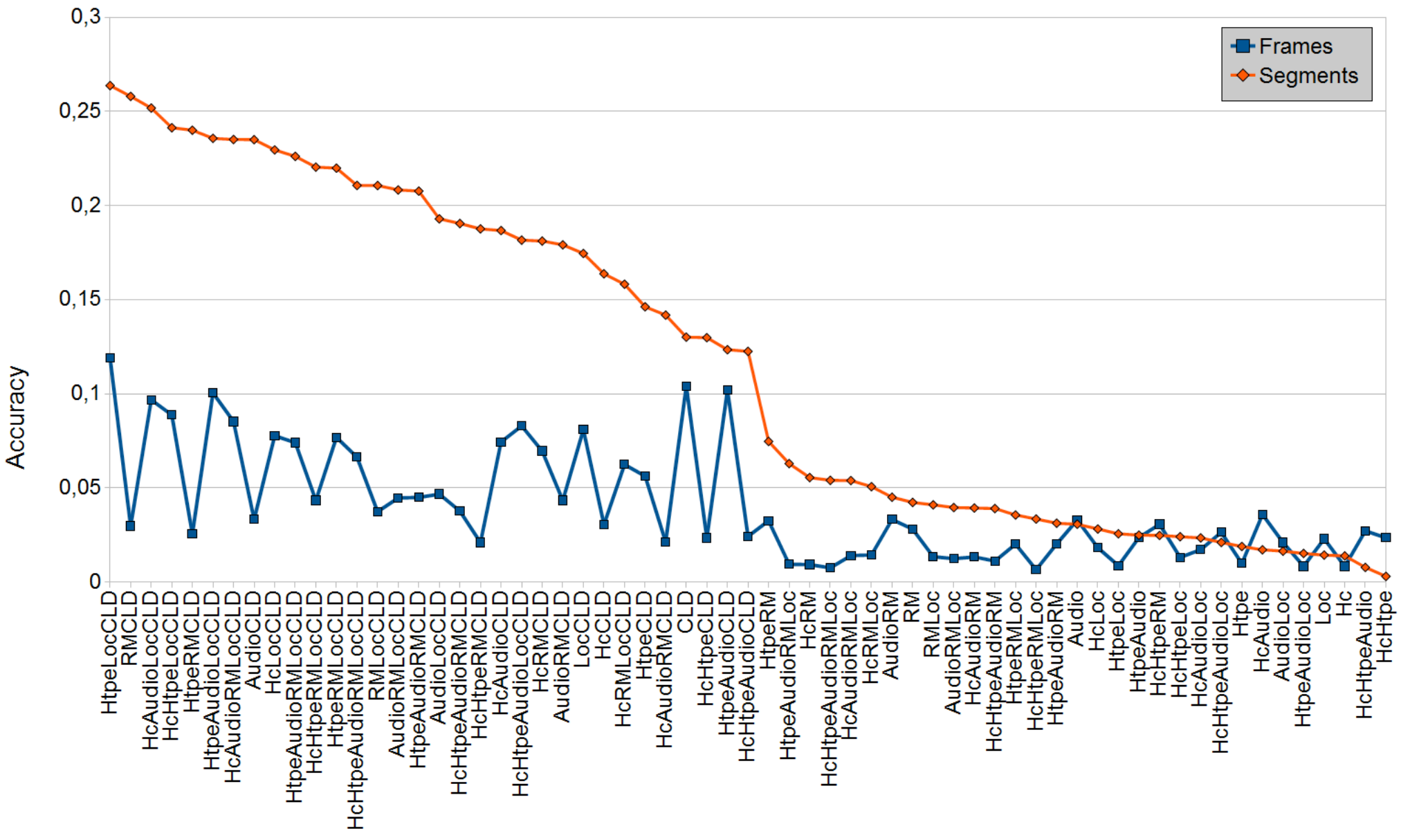

FIGURE 9: Global accuracy evaluation of recognition using frames (blue curve and square points) and segments (red curve and diamond points) over all the description spaces fusion tested (sorted by decreasing accuracy with respect to segments approach). Please refer to Table 1 for descriptors nomenclature.

NB: For a better readability of the figure, results are shown for a selected configuration (3statesNone1State) of the HMMs but are similar for other configurations.

each state, all observation models are initialized with 5 Gaussians mixtures except when the reject class “None” is modeled by one state which is then modeled by only one Gaussian.

The activities of interest considered for the first experiment correspond to the following ADLs: “Plant Spraying”, “Remove Dishes”, “Wipe Dishes”, “Meds Management”, “Hand Cleaning”, “Brushing Teeth”, “Washing Dishes”, “Sweeping”, “Making Coffee”, “Making Snack”, “Picking Up Dust”, “Put Sweep Back”, “Hair Brushing”, “Serving”, “Phone” and “TV”. We will specify a more complete list for the subsequent experiments. In the following we report the global performances of the evaluation for all activities and also more specific performances for some activities of interest. In the plots shown, the results per descriptor are sorted in decreasing order of performance.

### d. Evaluation of the influence of temporal segmentation

The proposed temporal segmentation reveals three main advantages. First, the amount of data to process in the recognition process is divided by a factor between 50 and 80 since one observation is defined for a segment and not for a frame. Second, the key frames may be used as a summary of the whole video which is relevant as it gathers the evolution of the patient in successive places. Finally, the evaluation of recognition performance presented in FIGURE 9 shows that the results are better when the recognition process is run on segments. In this figure the results are sorted in decreasing order. The best results are always obtained with segments as observations and other results are similar using frames or segments.

### e. Global evaluation of the description space

FIGURE 9 also shows which configurations are the most successful for the task. All the 33 best configurations are actually all the configurations including the $CLD$ descriptor. We will therefore in the following only consider configurations which include $CLD$, and evaluate all possible combinations of it with the other descriptors. The results are presented in FIGURE 10. Once again, a significant gain in performance can be observed when using segments instead of frames observations, the best accuracy for segments is 0.31 while the best accuracy for frames is 0.17. Here, the best global performance is obtained for the fusion $AudioCLD$ and good performances are also obtained for description spaces $RMCLD$, $H_cLocCLD$, $H_cAudioLocCLD$ and $H_cAudioRMLocCLD$. The $Audio$ descriptor seems efficient to capture some of the characteristic noises of activities which may occur for "Washing Dishes" or "Brushing Teeth" for example.

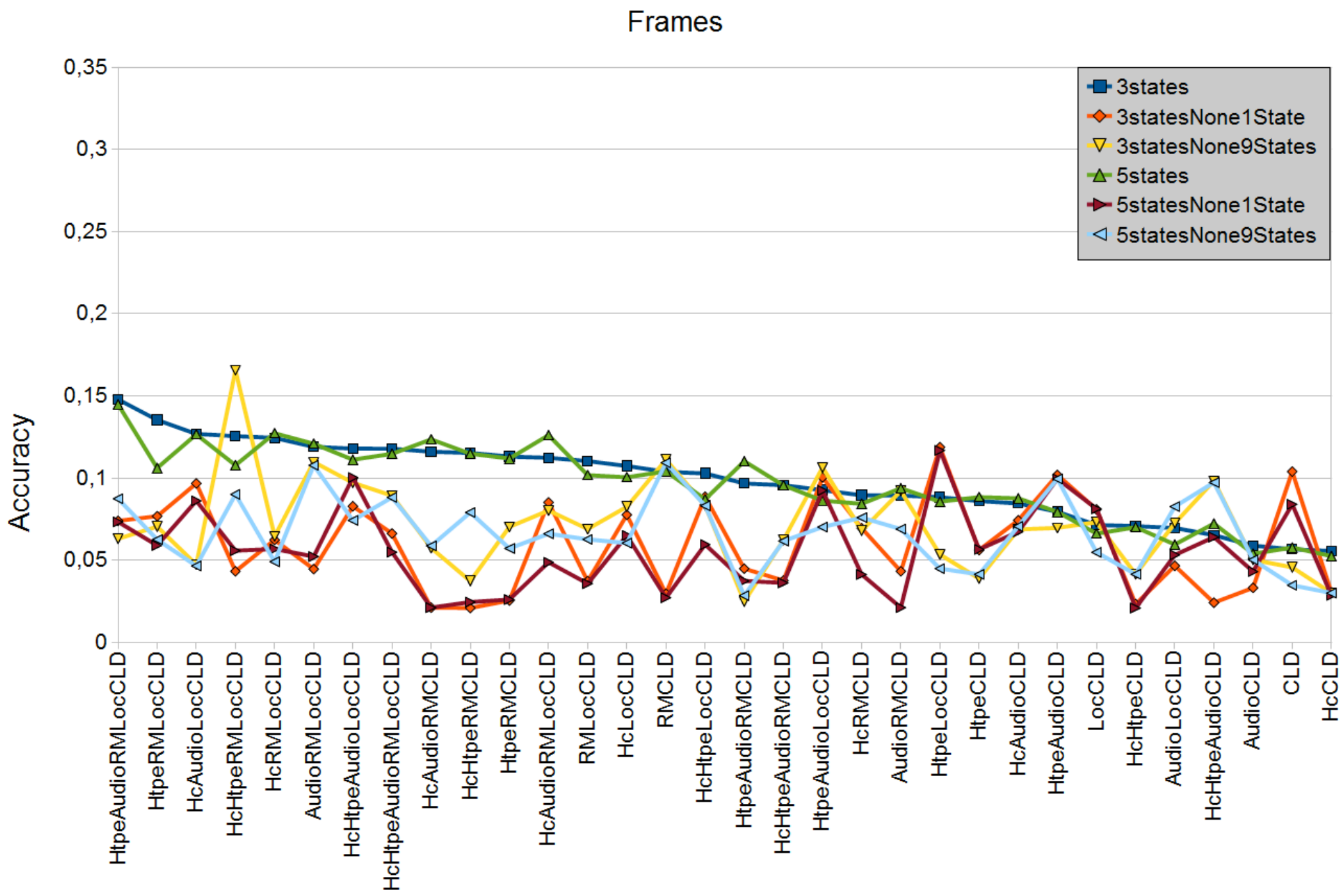

(a) Results using frames as observations

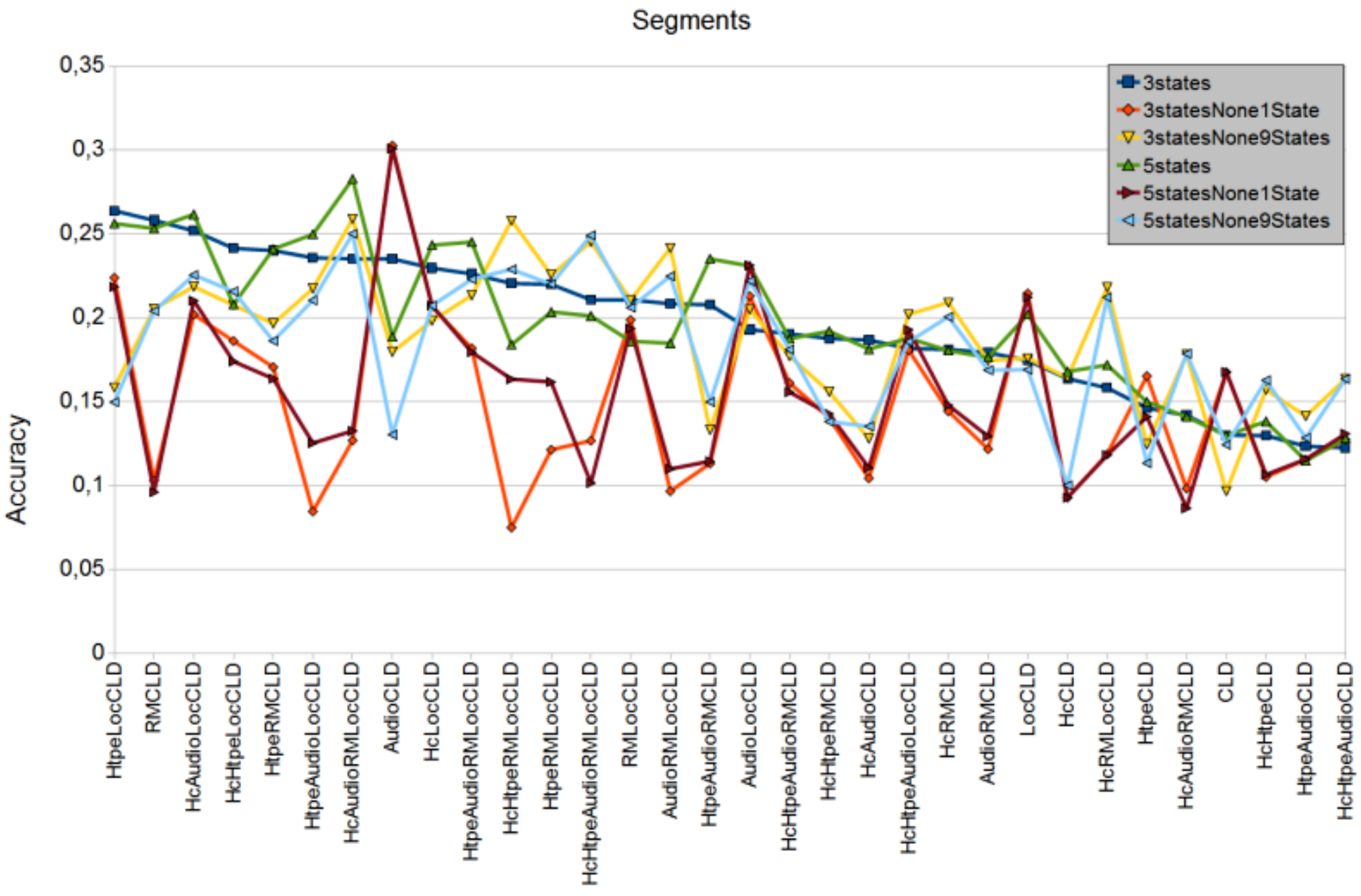

(b) Results using segments as observations

FIGURE 10: Global accuracy evaluation of recognition using segments over CLD and all possible fusion with CLD description spaces using frames (a) or segments (b) as observations. The curves represent 6 different HMM configuration: 3 states, 3 states with “None” class being modeled with only one state, 3 states with “None” class being modeled with 9 states, 5 states, 5 states with “None” class being modeled with only one state, 5 states with “None” class being modeled with 9 states.

### f. Global evaluation of the reject class model

We have also investigated the influence of modeling the reject class "None" in a different way than all the ADL classes. We have performed experiments when modeling this "None" class by a single state HMM or by a much more complex 9 states HMM. From the same FIGURE 10, we can see that performances with the reject class being modeled as a single state are clearly poorer and using 9 states does not significantly improve or degrade the performance. However, this configuration with 9 states for the "None" class shows good performances in high dimensionality description spaces built upon video segments.

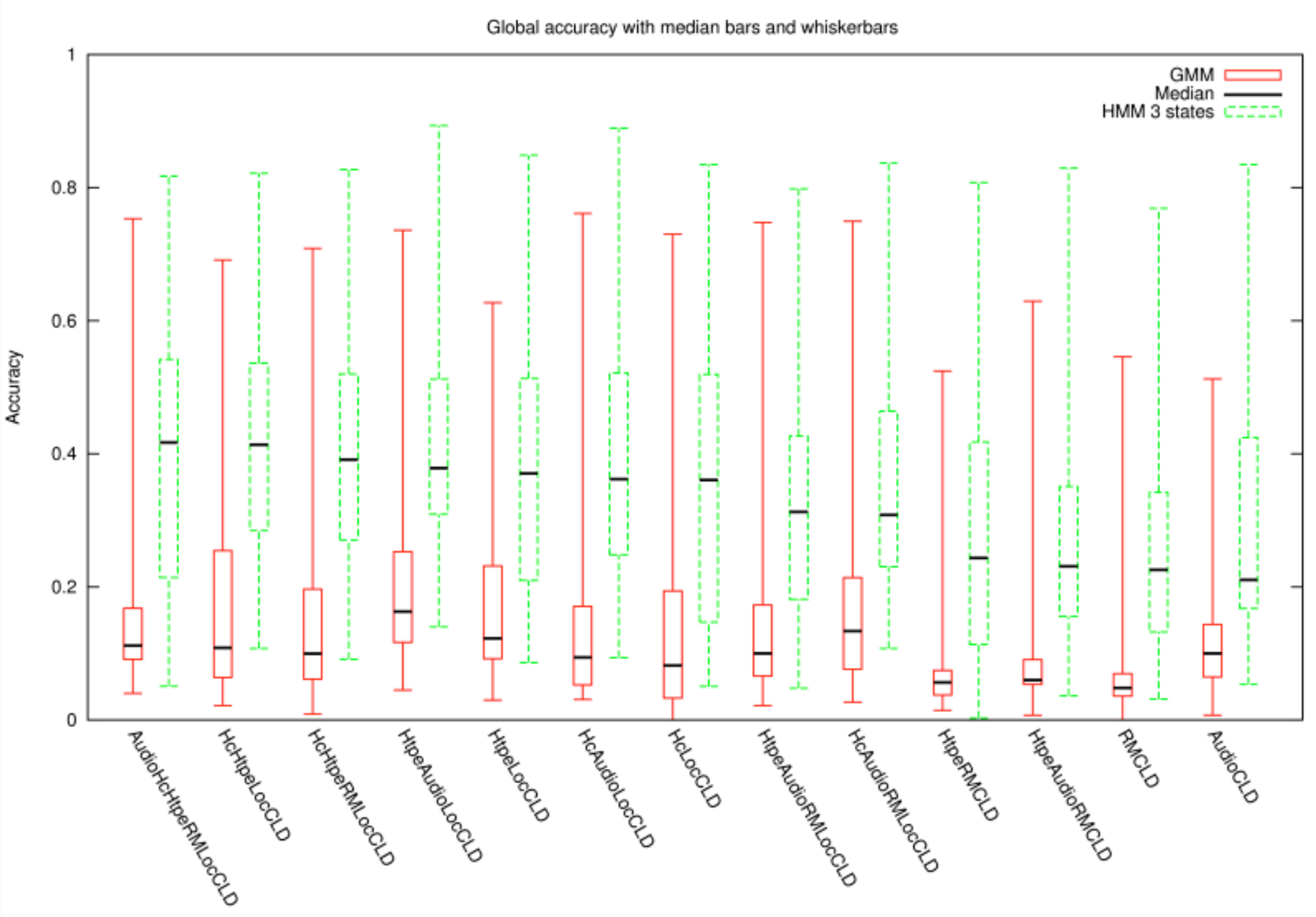

FIGURE 11: Global accuracy of the proposed approach and the GMM baseline with regards to all description space candidates. The results are sorted by decreasing median accuracy of the proposed approach.

### g. Evaluation recognition of activities on the whole corpus

Finally, the second experiment is run on a corpus of 26 videos following the same leave-one-out cross validation scheme. The description space candidates are the 13 best configurations from the first experiment. The number of states in the bottom-level HMM is fixed to 3. We have defined a GMM baseline where a GMM is learned for each activity on the same set of training data used for the HHMM approach. In this experiment, the 23 different activities are “Food manual preparation”, “Displacement free”, “Hoovering”, “Sweeping”, “Cleaning”, “Making a bed”, “Using dustpan”, “Throwing into the dustbin”, “Cleaning dishes by hand”, “Body hygiene”, “Hygiene beauty”, “Getting dressed”, “Gardening”, “Reading”, “Watching TV”, “Working on computer”, “Making coffee”, “Cooking”, “Using washing machine”, “Using microwave”, “Taking medicines”, “Using phone”, “Making home visit”.

An overview of the results in this larger scale experiments are given in FIGURE 11. The proposed approach clearly outperforms the GMM baseline. The best median accuracy for the GMM is 0.16, obtained for the description space $H_{tpe}AudioLocCLD$, while the HHMM approach obtains a best median accuracy of 0.42 for the complete description space $AudioH_cH_{tpe}RMLocCLD$. The gain of performance compared to the first experiment can be explained by the larger amount of training data. However, it is important to state the large variance of accuracy between 0.1 and 0.9. This shows the difficulty of our task.

### h. Evaluation for specific activities

A more in depth analysis of the performances for activities recognition is given in FIGURE 12. We have selected a subset of four activities (“Hoovering”, “Making a bed”, “Reading” and “Working on computer”) where the performances vary strongly when different description spaces are used. The performances are given in terms of accuracy, recall, precision and F-score. Note that the accuracy metric, when computed by activity, can easily be high since true negatives have positive impact on the performance. The results are sorted by decreasing precision as exchanges with the doctors have led to the conclusion that it was better to have less but more accurate detections, which is exactly what good precision metric values represent.

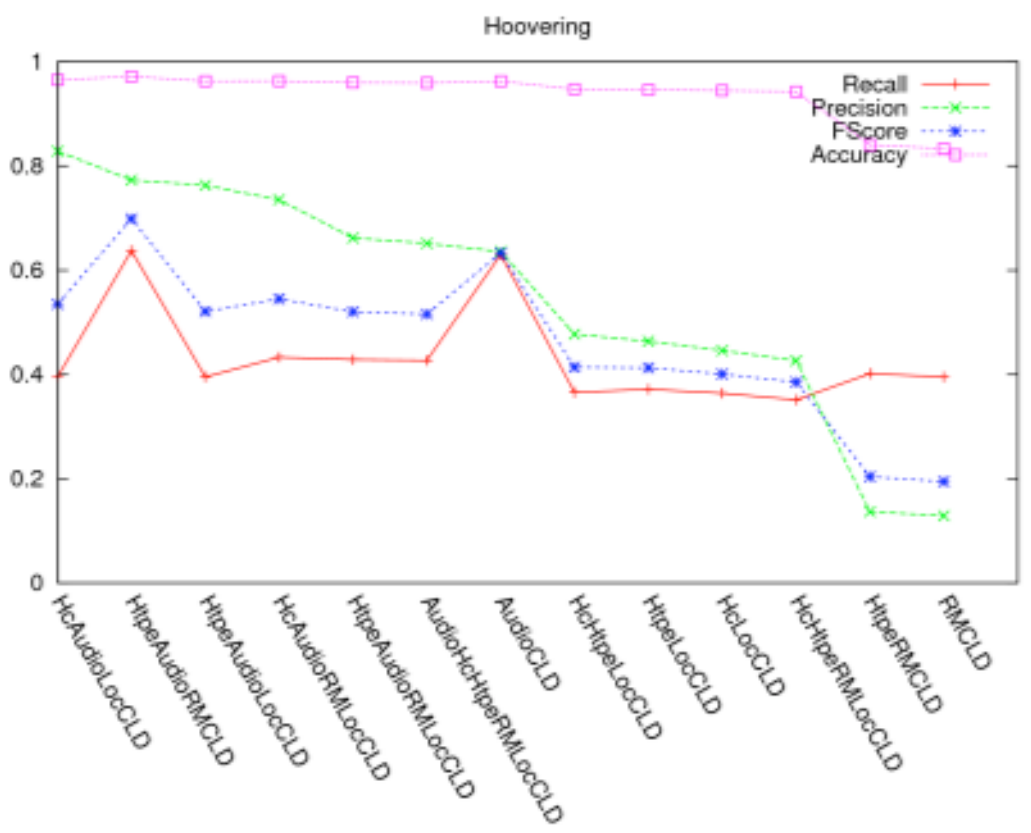

(a)

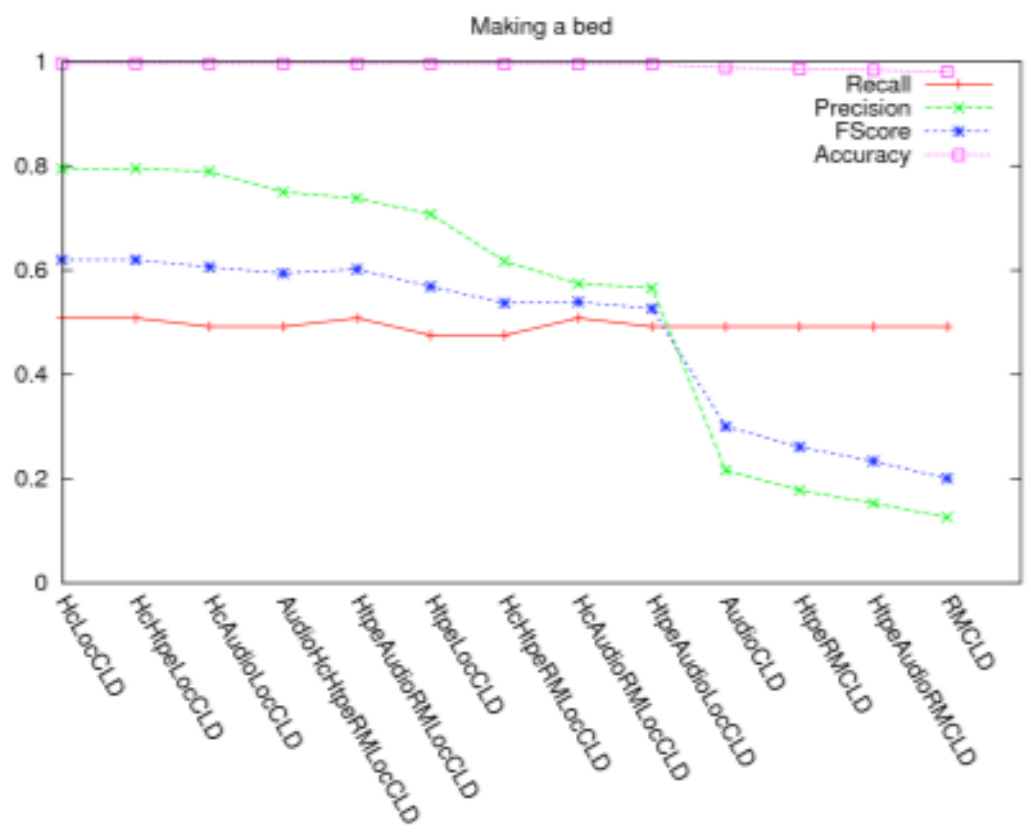

(b)

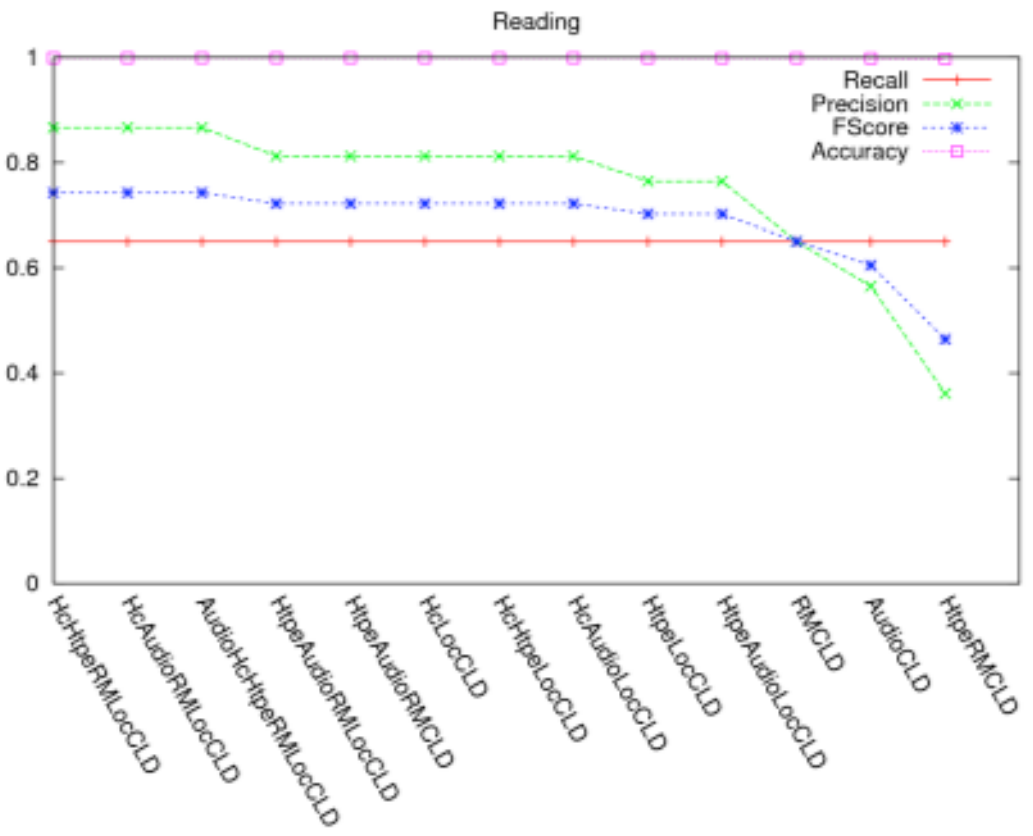

(c)

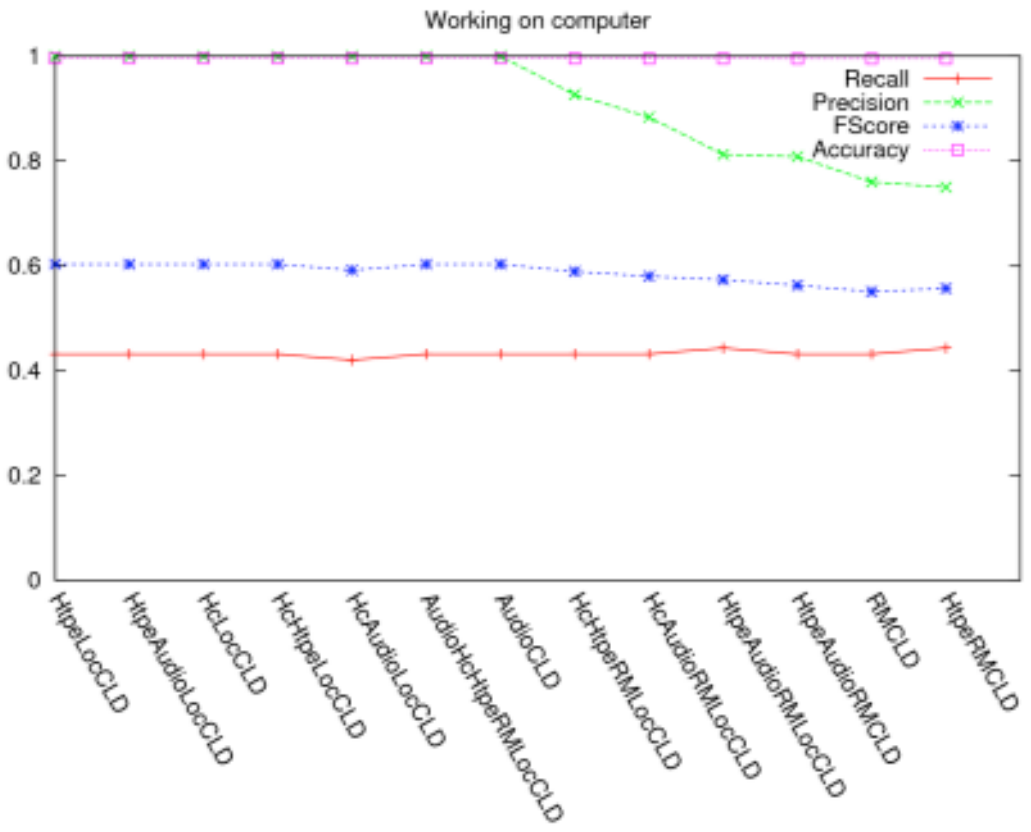

(d)

FIGURE 12: Performances according to the four metrics: recall (red curves, "+" points), precision (green curves, "x" points), F-score (blue curves, "*"points) and accuracy (pink curves, square points). The results are sorted by decreasing precision. a) "Hoovering", b) "Making a bed", c) "Reading" and d) "Working on computer".

For the activity "Hoovering", $Audio$ is essential as the 7 best performances contains the $Audio$ descriptor, see FIGURE 12.a. This can be clearly linked with the fact that one of the coefficients of the $Audio$ descriptor corresponds to the detection of a "hoover" sound. The best trade-off between recall and precision, i.e. the best F-score, is obtained for the description space $H_{tpe}AudioRMCLD$ which contains global and local instantaneous motion descriptors in addition to $Audio$ and $CLD$. The complete description space $AudioH_cH_{tpe}RMLocCLD$ is also a good trade-off between recall and precision.

The second activity studied is "Making a bed", the results are presented in FIGURE 12.b. For this activity the four top results contains the three descriptors $H_c$, $Loc$ and $CLD$, the best results being obtained for the description space $H_cLocCLD$. This activity will always happen in the bedroom, thus the presence of the $Loc$ descriptor in the best description spaces is not surprising. The $H_c$ descriptor is helpful to characterize the fact that a patient moves around the bed while performing this activity.

The third activity “Reading” is more static and involves localized residual motion while turning the pages of the book being read. This is confirmed as the 5 top description spaces incorporate the $RM$ descriptor, see FIGURE 12.c. The static component of the activity is captured by the motion descriptors. Both $H_c$ and $H_{tpe}$ seem efficient at capturing this characteristic. The activity “Reading” being more likely in a limited set of locations, the $Loc$ descriptor is also present in most of the best configurations.

FIGURE 12.d depicts the results for the last activity we studied: “Working on computer”. The best description spaces contain the $Loc$ and $CLD$ descriptors combined with at least one motion descriptor. Once again, the complete description space gives one of the best performances.

# 8. Conclusions and perspectives

In this paper we have tackled the problem of activity recognition in videos acquired with a camera worn by patients for the study of the dementia disease. These videos are complex, with strong and irregular motion and lighting changes, the presence of activities of interest in the recordings is rare.

We have proposed to solve the problem using the HMM formalism. A Hierarchical two level HMM was proposed to model both the semantic activities from the taxonomy defined by medical doctors and non-semantic intermediate states. In order to define the observations of the Hierarchical HMM we introduced a new concept of camera “viewpoint” and proposed a temporal segmentation of the video thanks to the analysis of the apparent motion.

We defined multimodal description spaces comprising motion features, static visual features and audio descriptors. Both low-level descriptors and mid-level features were used, with the objective to extract complementary and relevant information. The observations for training and recognition with the HHMM were obtained by the combination of the proposed features in an early fusion manner. We also introduced a state modeling the rejection class within the upper-level HMM. This is necessary for the description of our natural content, which contains transitions between the activities of interest and non-relevant actions of the patients.

The proposed model was tested on the unique-in-the-world video corpus acquired with healthy volunteers and patients in an "ecological" environment, i.e. at their homes. The taxonomy of activities was defined by medical researchers and the proposed framework was tested with cross-validation to recognize them. In these tests the optimal configurations of description space ensure performance which is nearly 8 times better than chance and 4 times better than a GMM baseline.

Detailed studies of different description spaces and HHMM states configurations have revealed that:

- (i) temporal segmentation into view-points improves the performances due to the filtering of the descriptors within meaningful units of time;
- (ii) for the bottom-level HMM in our hierarchical model, three states are sufficient to model the internal structure of each semantic activity;
- (iii) as far as the description space is concerned, the complete description space combining all available features performs best in terms of median accuracy. Even when it is difficult to choose an absolute winner for description space composition, the best overall performances are ensured when static color descriptors of a scene content are present;
- (iv) the optimal description space varies per activity, each descriptor bringing more information for a specific activity; this often correlates with "common sense": e.g. the hoovering activity is the best recognized with audio features in description space.

The last statement makes us think that despite interesting performances the proposed framework reaches for this challenging application, the future is in the incorporation of more semantic features in the description space. Events from more complete wearable sensor sets can be used, such as accelerometers and others. The combination of such sensors with wearable video and audio offers new avenues to be explored. In video, we think about defining a concept flow related to the recognition of objects that the person manipulates. This would leverage the fusion of information from video and other sources. Last, the good acceptance of the wearable sensor device by patients, indicate that the proposed approach has the potential for direct clinical perspectives.

## Acknowledgements

This work is partly supported by a grant from the ANR (Agence Nationale de la Recherche) with reference ANR-09-BLAN-0165-02, within the IMMED project.

# 9. References

[1] H. Amieva, M. Le Goff, X. Millet, J.-M. Orgogozo, K. Pérès, P. Barberger-Gateau, H. Jacqmin-Gadda and J.-F. Dartigues, "Prodromal Alzheimer's disease: Successive emergence of the clinical symptoms", *Annals of Neurology*, volume 64, issue 5, pages 492–498, November 2008.
[2] N. Zouba, F. Bremond, A. Anfonso, M. Thonnat, E. Pascual, and O. Guerin, "Monitoring elderly activities at home", *Gerontechnology*, volume 9, issue 2, May, 2010.
[3] R. Hamid, S. Maddi, A. Johnson, A. Bobick, I. Essa and Ch. Isbell, "A novel sequence representation for unsupervised analysis of human activities", *Artificial Intelligence* (2009), volume 173, pages 1221–1244.
[4] R. Megret, D. Szolgay, J. Benois-Pineau, P. Joly, J. Pinquier, J.-F. Dartigues and C. Helmer, "Wearable video monitoring of people with age dementia: Video indexing at the service of healthcare", *International Workshop on Content-Based Multimedia Indexing - CBMI* (2008), Conference Proceedings, art. no. 4564934, pages 101-108.
[5] S. Hodges, L. Williams, E. Berry, S. Izadi, J. Srinivasan, A. Butler, G. Smyth, N. Kapur and K. R. Wood,, "Sensecam: a retrospective memory aid". *UBICOMP'2006*, pages 177–193, 2006.
[6] L. Piccardi, B. Noris, O. Barbey, A. Billard, G. Schiavone, F. Keller and C. von Hofsten. "Wearcam: A head wireless camera for monitoring gaze attention and for the diagnosis of developmental disorders in young children". *International Symposium on Robot & Human Interactive Communication*, pages 177-193, 2007.
[7] HTK Web-Site: http://htk.eng.cam.ac.uk
[8] L. R. Rabiner, "A tutorial on hidden markov models and selected applications in speech recognition". *Proceedings of the IEEE*, volume 77, number 2, pages 257-286, 1989.
[9] J. S. Boreczky and L. D. Wilcox, "A Hidden Markov Model Framework for Video Segmentation Using Audio and Image Features". Proceedings of the IEEE International Conference on *Acoustics, Speech and Signal Processing*, volume 6, pages 3741-3744, 1998.
[10] E. Kijak, G. Gravier, P. Gros, L. Oisel and F. Bimbot, "HMM based structuring of tennis videos using visual and audio cues" *International Conference on Multimedia and Expo* – ICME (2003), volume 3, pages 309-312.
[11] S. P. Chatzis, D. I. Kosmopoulos, and T. A. Varvarigou, "Robust sequential data modeling using an outlier tolerant hidden markov model". *IEEE Transactions on Pattern Analysis and Machine Intelligence*(2009), 31 (9), pages 1657-1669.
[12] Y. Ivanov, and A. Bobick, "Recognition of visual activities and interactions by stochastic parsing". *IEEE Transactions on Pattern Analysis and Machine Intelligence*, 22 (8), pages 852-872, August 2000.
[13] S. Fine, Y. Singer and N. Tishby "The Hierarchical Hidden Markov Model: Analysis and Applications". *Machine learning* (1998), volume 32, pages 41-62.
[14] M. Gales and J. Young "The Theory of Segmental Hidden Markov Models". *University of Cambridge*, Department of Engineering, 1993.
[15] M. Ostendorf, V. Digalakis and O. A. Kimball, "From HMMs to Segment Models: A Unified View of Stochastic Modeling for Speech Recognition", *IEEE Transactions on Speech and Audio Processing* (1995), volume 4, pages 360-378.
[16] M. Delakis, G. Gravier and P. Gros, "Audiovisual integration with Segment Models for tennis video parsing" *Computer Vision and Image Understanding* (2008), volume 111, number 2, pages 142-154.

[17] D. Surie, T. Pederson, F. Lagriffoul, L-E. Janlert and D. Sjölie, "Activity Recognition using an Egocentric Perspective of Everyday Objects". *Ubiquitous Intelligence and Computing* (2007), Springer, pages 246-257.
[18] J. Benois-Pineau and P. Kramer, "Camera motion detection in the rough indexing paradigm". *TREC Video*, 2005.
[19] S. Lazebnik, C. Schmid and J. Ponce, "Beyond bags of features: Spatial pyramid matching for recognizing natural scene categories". *IEEE Conference on Computer Vision and Pattern Recognition* - CVPR, (2006) volume 2, pages 2169-2178.
[20] F. Jurie and B. Triggs, "Creating efficient codebooks for visual recognition", *Tenth IEEE International Conference on Computer Vision - ICCV* (2005), volume 1, pages 604-610.
[21] B. Scholkopf, A. Smola and K.R. Muller, "Nonlinear component analysis as a kernel eigenvalue problem", *Neural computation* (1998), volume 10 (6), pages 1299-1319.
[22] H. Bay, T. Tuytelaars and Luc Van Gool, "SURF: speeded-up robust features", *Computer Vision and Image Understanding* (2008), volume 110 (3), pages 346-359.
[23] Y. Bengio, O. Delalleau, N. Le Roux, J.-F. Paiement, P. Vincent and M. Ouimet, "Spectral dimensionality reduction", *Feature Extraction* (2006), Foundations and Applications, Springer, pages 519-550.
[24] C. Burges, "A tutorial on support vector machines for pattern recognition", *Data mining and knowledge discovery* (1998), volume 2 (2), pages 121-167.
[25] R. André-Obrecht, "A new statistical approach for automatic speech segmentation". *IEEE Transactions on Audio, Speech and Signal Processing* (1988), volume 36(1), pages 29–40.
[26] J. Pinquier and R. André-Obrecht, "Audio indexing: Primary components retrieval - robust classification in audio documents". *Multimedia Tools and Applications* (2006), volume 30(3), pages 313–330.
[27] G. Quenot, J. Benois-Pineau, B. Mansencal, E.Rossi, et al. ., "Rushes summarization by IRIM consortium: redundancy removal and multi-feature fusion". *VS'08 (Trec Video Summarization)*, 2008.
[28] T. Sikora, B. Manjunath, and P. Salembier, "Introduction to MPEG-7: Multimedia content description interface". 2002.
[29] Y. Gaëstel, C. Onifade-Fagbemi, F. Trophy, S. Karaman, J. Benois-Pineau, R. Mégret, J. Pinquier, R. André-Obrecht and J.-F. Dartigues "Autonomy at home and early diagnosis in Alzheimer Disease: usefulness of video indexing applied to clinical issues. The IMMED Project". *Alzheimer's Association International Conference on Alzheimer's Disease - AAICAD*, 16-21 Juillet, 2011, France
[30] S. J. Young and S. Young, "The HTK hidden Markov model toolkit: Design and philosophy". *Entropic Cambridge Research Laboratory, Ltd*, 1994.
[31] S. Young, G. Evermann et al., "The HTK book". 1997.
[32] S. Karaman, J. Benois-Pineau, J.-F. Dartigues, Y. Gaëstel, R. Mégret and J. Pinquier, "Activities of Daily Living Indexing by Hierarchical HMM for Dementia Diagnostics". *Content-Based Multimedia Indexing and retrieval - CBMI'2011*, IEEE Workshop, 13-15 Juin, 2011, Madrid, Espagne.
[33] N. Harte, D. Lennon, and A. Kokaram, "On parsing visual sequences with the hidden Markov model". *EURASIP Journal on Image and Video Processing*, pages 1-13, 2009.
[34] P. Guyot, J. Pinquier and R. André-Obrecht. "Water flow detection from a wearable device with an new feature, the spectral cover", submitted to CBMI'2012, IEEE Workshop, 27-29 June, 2012, Annecy, France.
[35] S. Galliano, E. Geofrois, De. Mosterfa, J.F. Bonastre, and G. Gravier, "The Ester phase II evaluation campaign for the rich transcription of the French broadcast news," EUROSPEECH, pp. 1149–1152, 2005.
[36] E.H. Spriggs, F. De La Torre and M. Hebert. "Temporal Segmentation and Activity Classification from First-person Sensing." In First *Workshop on Egocentric Vision*, 2009, pages 17-24.
[37] S. Sundaram and W. Mayol-Cuevas. "High Level Activity Recognition using Low Resolution Wearable Vision". In First Workshop on Egocentric Vision, 2009, pages 25-32.
[38] S. Sundaram and W. Mayol-Cuevas. "Egocentric Visual Event Classification with Location-Based Priors." In International Symposium on Visual Computing, 2010, Lecture Notes in Computer Science volume 6454, pages 596-605.
[39] J. Liu, J. Luo and M. Shah. "Recognizing realistic actions from videos 'in the wild'." In IEEE Conference on *Computer Vision and Pattern Recognition*, 2009, pp. 1996-2003.
[40] First Workshop on Egocentric Vision, held in conjunction with CVPR 2009.
[41] D. Byrne, A.R. Doherty, G.J.F Jones, A.F Smeaton, S. Kumpulainen and K. Järvelin. "The SenseCam as a tool for task observation." In Proceedings of the *22nd British CHI Group Annual*

*Conference on HCI* 2008: People and Computers XXII: Culture, Creativity, Interaction-Volume 2, 19–22, 2008.
[42] A. Doherty, N. Caprani, C. Ó Conaire, V. Kalnikaite, C. Gurrin, A. F. Smeaton and N. E. O'Connor, "Passively recognising human activities through lifelogging". *Computers in Human Behavior*, Volume 27, Issue 5, September 2011, pages 1948–1958.
[43] C. Schuldt, I. Laptev and B. Caputo, "Recognizing human actions: A local SVM approach", Proceedings of the 17th *International Conference on Pattern Recognition* (ICPR'2004), pp. 32-36, 2004.
[44] L. Ballan, M. Bertini, A. Del Bimbo, L. Seidenari and G. Serra, "Event Detection and Recognition for Semantic Annotation of Video", *Multimedia Tools and Applications*, number 1, volume 51, pp. 279-302, 2011.
[45] R. Poppe, "A survey on vision-based human action recognition", *Image and Vision Computing*, number 6, volume 28, pages 976-990, 2010.
[46] Z. Gao, M. Chen, A. Hauptmann, and A. Cai, "Comparing evaluation protocols on the KTH dataset", *International Conference on Human Behavior Understanding* - HBU, 2010, LNCS volume 6219, pp. 88-100.
[47] D. Gorisse, F. Precioso, P. Gosselin, L. Granjon, D. Pellerin, M. Rombaut, H. Bredin, L. Koenig, R. Vieux, B. Mansencal, J. Benois-Pineau, H. Boujut, C. Morand, H. Jégou, S. Ayache, B. Safadi, Y. Tong, F. Thollard, G. M. Quénot, M. Cord, A. Benoît and P. Lambert, "IRIM at TRECVID 2010: Semantic Indexing and Instance Search", *Proc. TRECVID 2010 Workshop*, 2010.
[48] M. Hill, G. Hua, A. Natsev, JR. Smith, L. Xie, B. Huang, M. Merler, H. Ouyang, M. Zhou, "IBM Research TRECVID-2010 Video Copy Detection and Multimedia Event Detection System", *Proc. TRECVID 2010 Workshop*, 2010.
[49] Z.-Z. Lan, L. Bao, S.-I Yu, W. Liu, A. G. Hauptmann, "Double Fusion for Multimedia Event Detection", *International Conference on Multimedia Modeling (MMM'12)*, pp. 173-185, 2012.

---

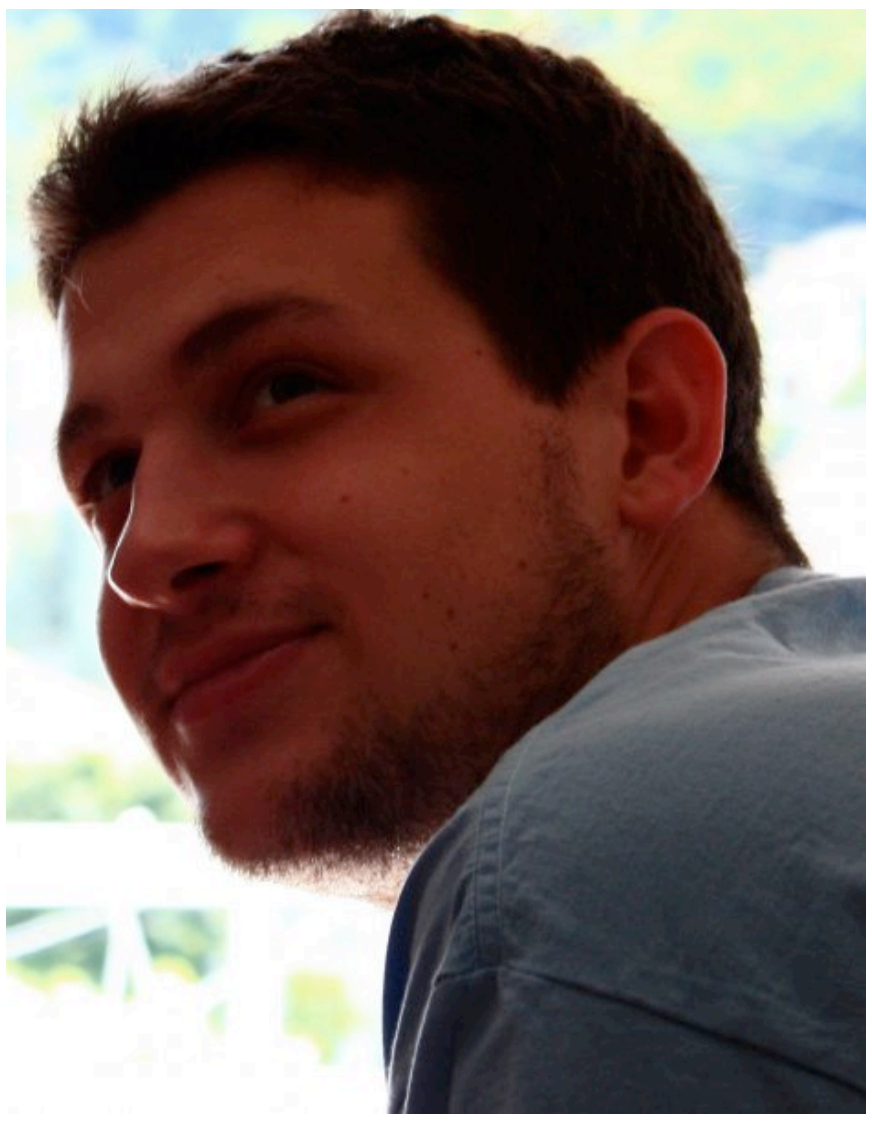

**Svebor Karaman** received a master's degree in computer engineering from the University of Bordeaux and an engineer diploma from the ENSEIRB in 2008. He has obtained a Ph.D. in Computer Science from the University of Bordeaux in 2011, with his thesis entitled "Indexing of Activities in Wearable Videos: Application to Epidemiological Studies of Aged Dementia". He has joined the MICC – Media Integration and Communication Center, at the beginning of 2012 as a postdoctoral researcher. His research interests focus on computer vision, semantic concepts recognition in images and videos, multimedia information retrieval and problematics related to the use wearable videos. During his PhD thesis, he has worked on human activities recognition by Hidden Markov Models (HMM) in videos recorded from a wearable device. He also proposed an object recognition approach in the Bag-of-Visual-Words framework which integrates spatial information within semi-local features: the Graph-Words.

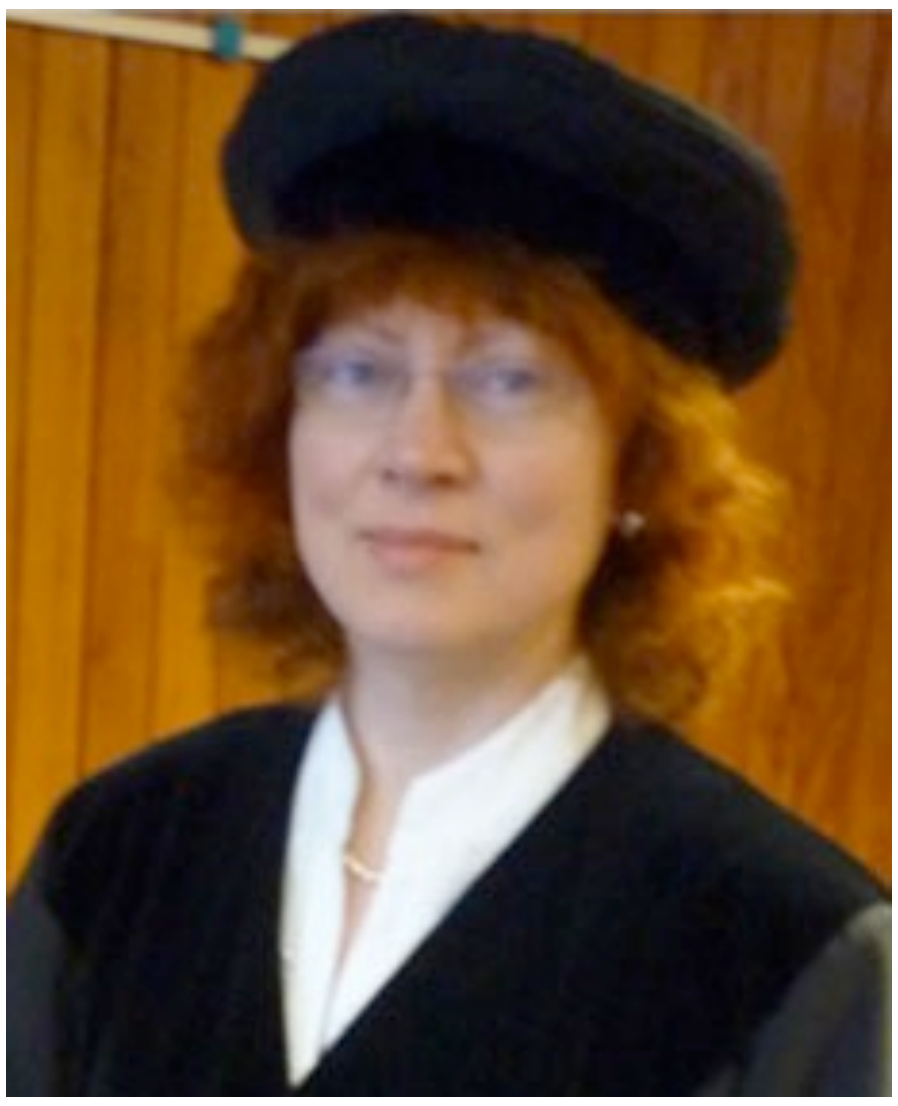

**Jenny Benois-Pineau** is a full professor of Computer science at the University Bordeaux 1 and chair of Video Analysis and Indexing research group in Image and Sound Department of LABRI UMR 58000 Université- Bordeaux 1 / Bordeaux 2 / CNRS / IPB-ENSEIRB-Matmeca. She is also a deputy scientific director of theme B of French national research unity GDR CNRS ISIS, and director of Computer Science Department of the faculty of Mathematics and Computer science at the University Bordeaux 1.

She is the author and co-author of more than 110 papers in international journals, conference proceedings, book chapters. She has tutored and co-tutored 18 PhD students and 19 research masters.

She is associated editor of EURASIP Signal Processing: Image Communication, Elsevier, Multimedia Tools and applications, Springer, TS Hermes-Lavoisier journals. She has served in numerous program committees in international conferences and workshops: ACM MM, CVIR, CBMI, AMR, IPTA, SAMT, ECMCS...

She has served as expert for European Commission since FP4. She is a member of Multimedia Commis- sion of French Ministry of National Education and member of scientific board of International Center for Mathematical Modelling at the University of Växjo, Sweden.

She has been coordinator or leading researcher in international research projects Platon, Balaton, IP XMedia, IP Dem@care, French representative in COST292 European action, national research projects ANR and numerous projects with French industrial companies. She gave invited lectures at the universities of Sussex (GB), UPC (Spain), UNAM (Mexico), University of North Carolina at Chapel Hill (USA), Brooklynn Polytechnic (USA), Firenze (Italy)...

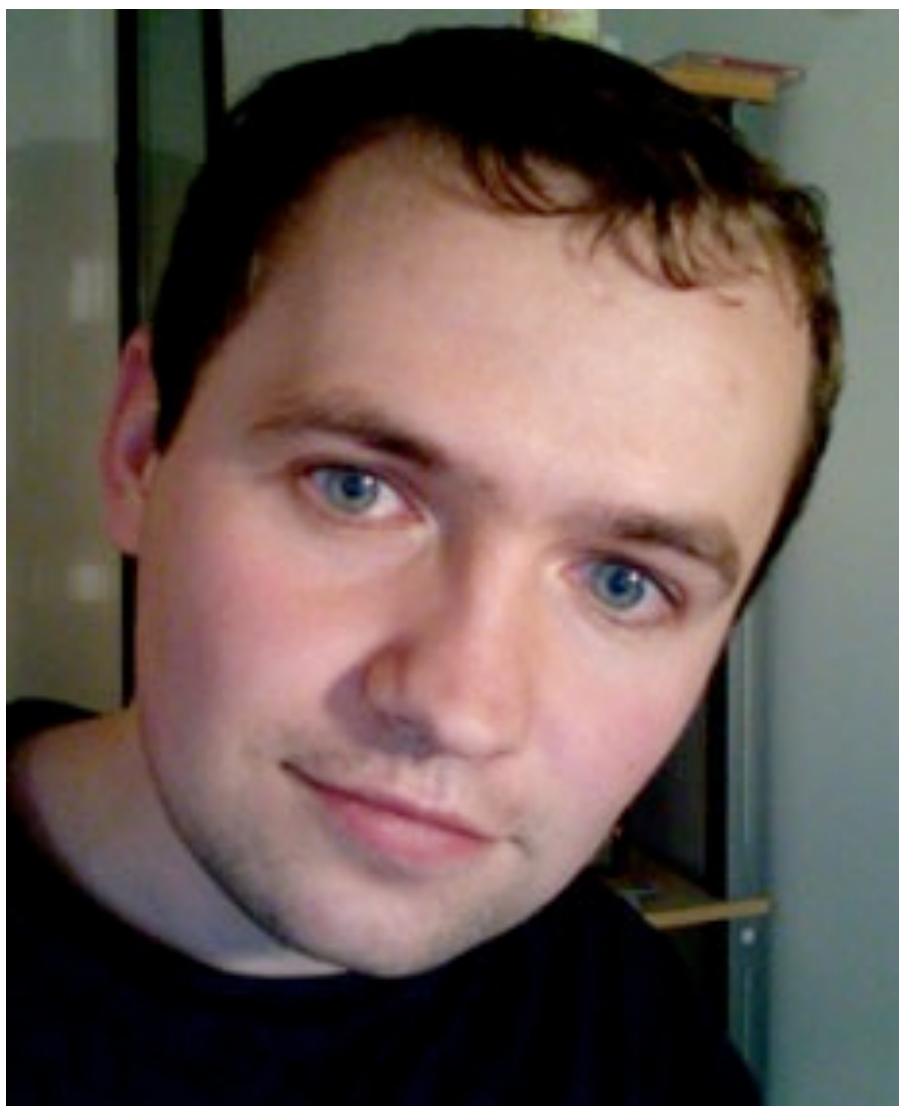

**Vladislavs Dovgalecs** received Master degree in Computer Science with specialization in mathematic fundamentals of computers science and satellite information data processing systems in 2008 from University College of Ventspils, Latvia. In the last year of master studies he joined INPL de Nancy, France as an ERASMUS student and successfully defended his master thesis. In 2011 he obtained Ph.D. in Computer Science from Signal and Image Processing Group at IMS laboratory, UMR 5218 CNRS, University of Bordeaux, France with involvement in the ANR project IMMED. He is currently postdoctoral research engineer with LITIS Laboratory, University of Rouen, France. He is involved in Interreg project Docexplore to develop advanced word spotting algorithms and tools for expert annotation and user search in ancient manuscripts.

His main areas of interest are multimedia indexing, pattern recognition and applied machine learning in computer vision.

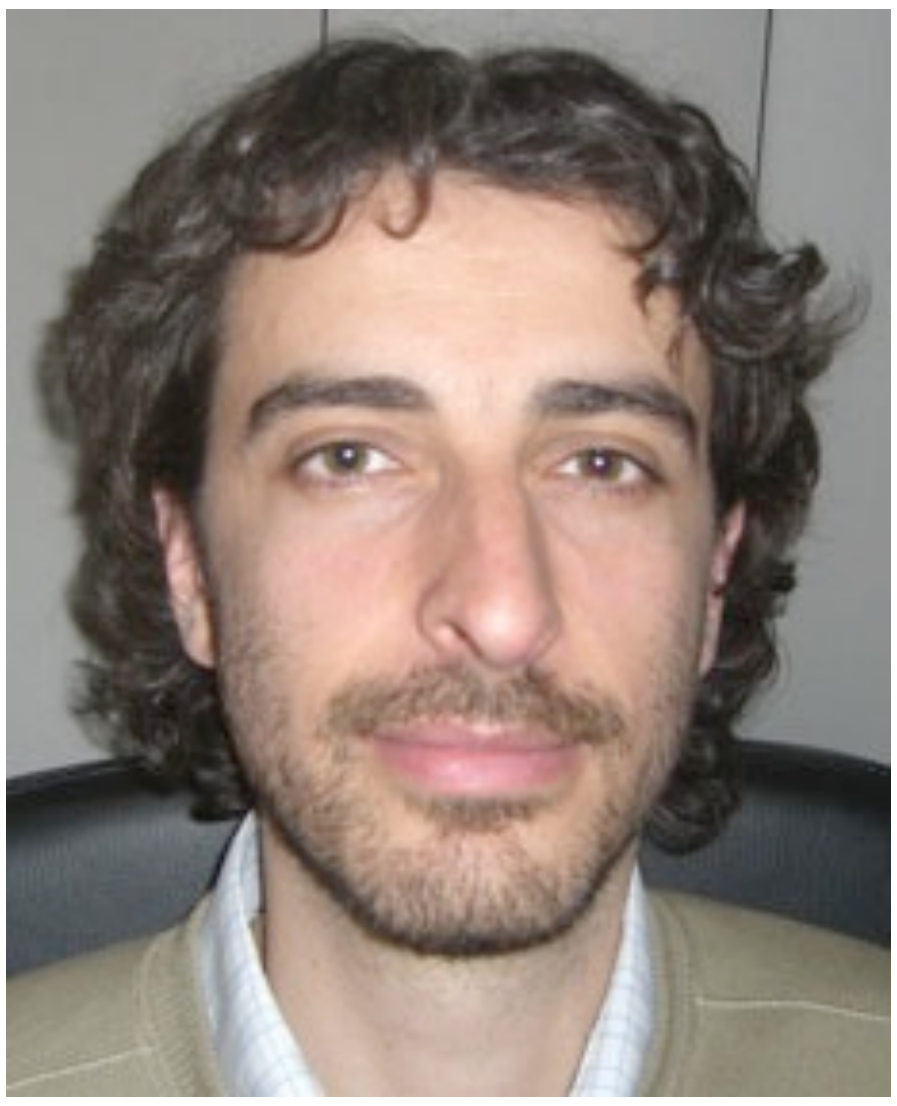

**Rémi Mégret** received the Magistère in Computer Science and Modeling in 2000 from Ecole Normale Supérieure de Lyon, France, and the Ph.D. in Computer Science from INSA de Lyon, France, in 2003. As a Ph.D. student, he was a member of the LIRIS Laboratory, INSA de Lyon, and a visiting student at the LAMP laboratory, University of Maryland. In 2004, he joined the Bordeaux Institute of Technology (ENSEIRB/IPB). He is currently associate professor within the Signal and Image Processing Group at IMS Laboratory, UMR 5218 CNRS, University of Bordeaux, France. His main areas of interest are video analysis, pattern recognition and computer vision. He is involved in the development of methods for activity analysis from video acquired with wearable cameras within the ANR project IMMED and the European project Dem@care.

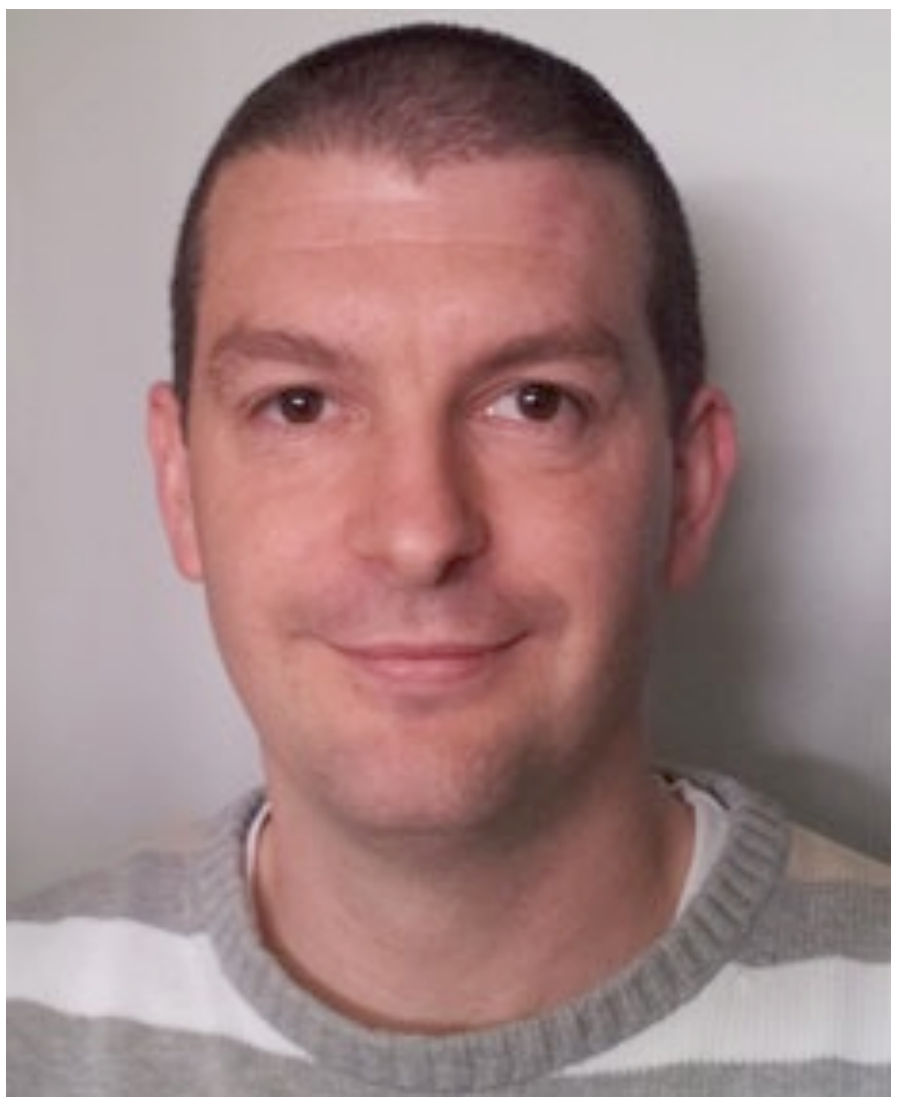

**Julien Pinquier** was born in 1977. He obtained a PhD (computer science speciality) in 2004. Its work concerned audio indexing and structuring by search of primary components: speech, music and keysounds (jingles, applause, keywords, etc.). Since 2005, he is an assistant professor at the Paul Sabatier University where he works in the IRIT laboratory. Its objectives relate to the combination of the audio and the video, the multimedia indexing for automatic structuring of audiovisual documents. Indeed, to have robust audio and video classification techniques and to take into account a large volume of data.

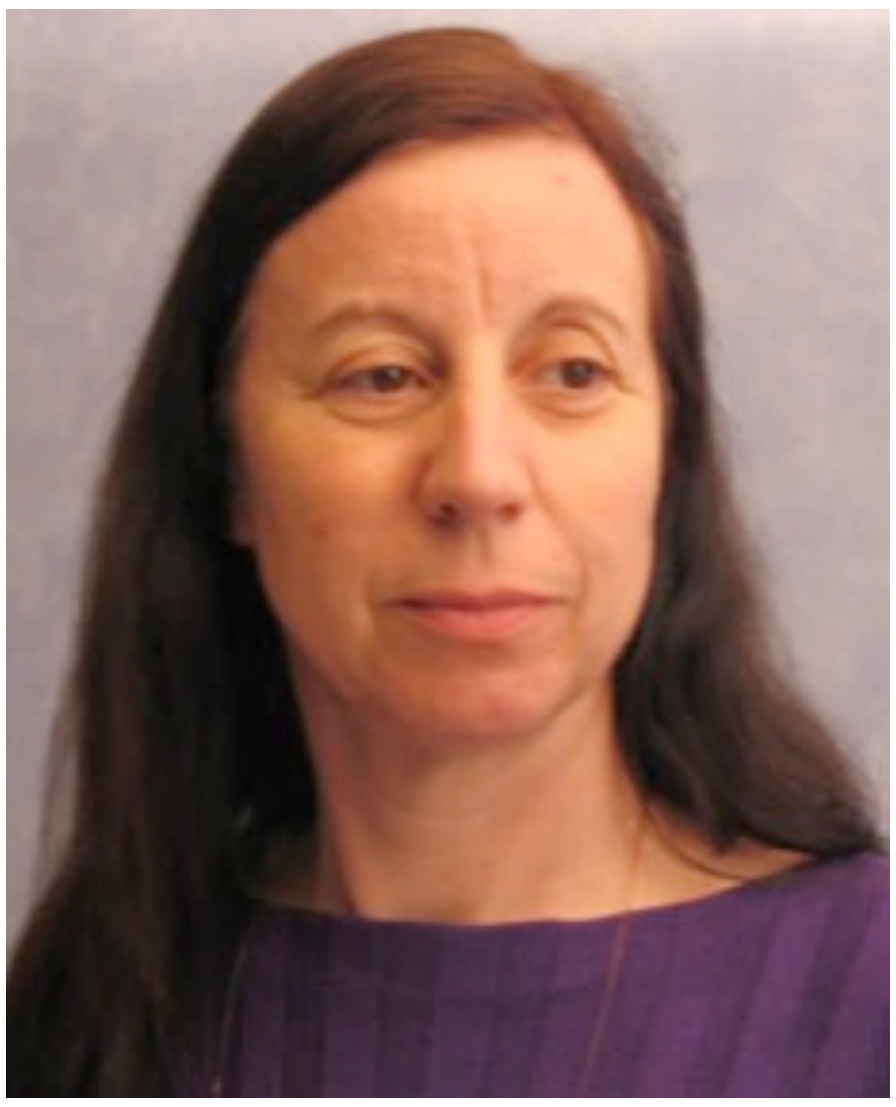

**Régine André-Obrecht** was born in Orléans (France) in 1955. She studied mathematics and probability theory at Ecole Normale Supérieure of Fontenay aux Roses, and she received the diploma of “Agregation de Mathematiques” in 1981. In 1982-1985, she worked with the Institut de Recherches en Informatique et Systèmes Aléatoires de Rennes (IRISA) at the department of Signal Processing with Albert Benveniste. She received the doctorat of 3e cycle from the University of Rennes in 1985. Since she is a researcher at the CNRS (National Center for Scientific Research): between 1986 and 1991, she worked at IRISA, and in 1991 she joined the Institut de Recherches en Informatique de Toulouse (IRIT) of the University Paul Sabatier of Toulouse. She received the diploma “Habilitation à

Diriger des Recherches" of the University of Rennes in 1993. She is professor of the University Paul Sabatier since 1999.

Her main research topics are Speech analysis (segmentation and robust analysis for speech recognition, information fusion), Speech, Speaker and Language recognition (audio visual recognition, sound duration modeling), audio content description and audio-video structuration (speech/music discrimination, jingle detection). Head of the group SAMoVA (Structuration, Analysis, Modeling of Video and Audio documents) since 2002 (7 permanent researchers and 8 PhD students), Régine André-Obrecht is implied in several scientific projects (national as ANR projects, European as the NoE Muscle). Member of the Direction Committee of the National GDR (Group of Research) (Information, Signal, Image, Vision), she was a main actor of the Multimedia Indexing French community; she co-organized the first CBMI workshop (1999) at Toulouse and she is a member of the steering committee of the CBMI workshop since its founding. Member of the International Speech Communication Association, presently, she is a member of the program committee of Interspeech 2007, RFIA 2008 and CARI 2008. Invited Editor for a special issue of Speech Communication (July 2000), she is a member of the Redaction Committee of the Journal Traitement de Signal and the Journal ARIMA.

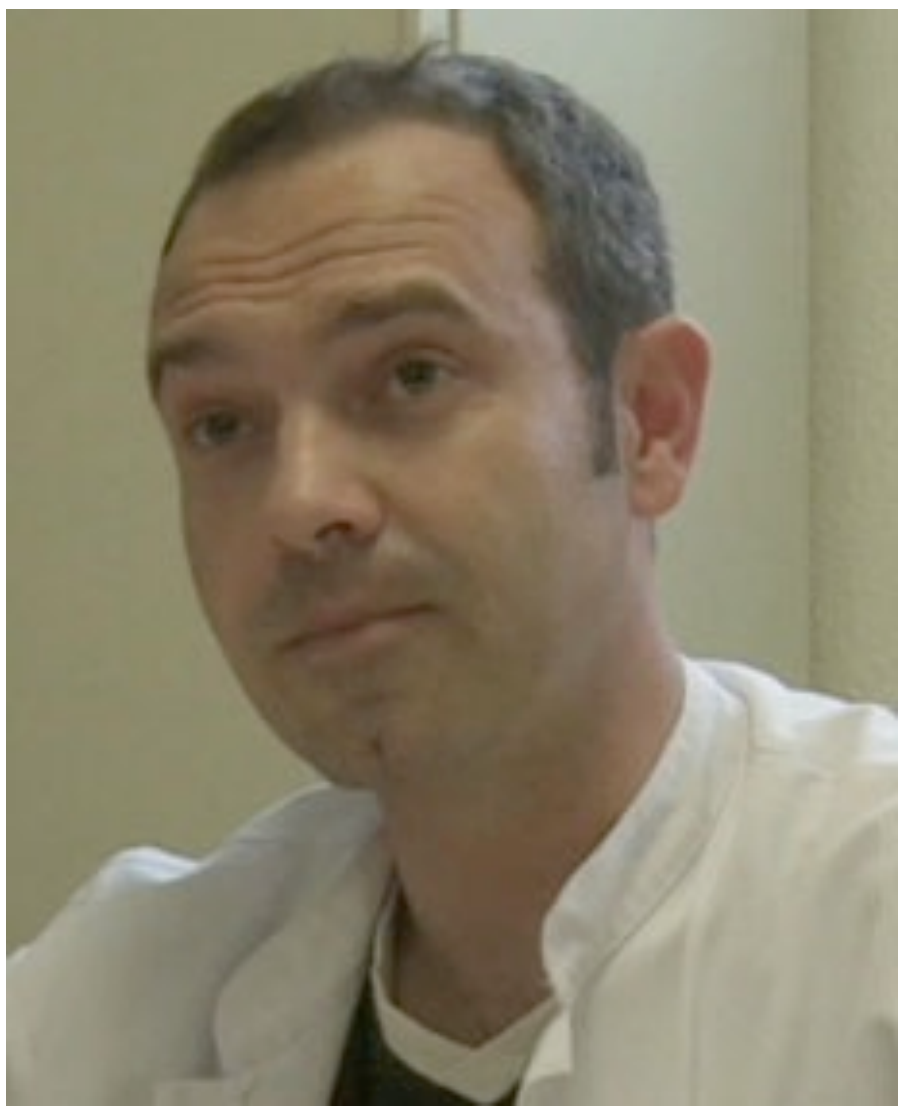

**Yann Gaëstel** got his PhD in Medical and Biological Sci from University Bordeaux Segalen in 2005. He is actually the research director at Center of Health Bagatelle. He has been a lecturer at the University Bordeaux Segalen and associated researcher at INSERM. Dr. Gaestel is psycho-motricien and neuroscientist, he is authors and co-authors of numerous research publications in the field of dementia studies and application of IT for this purpose.

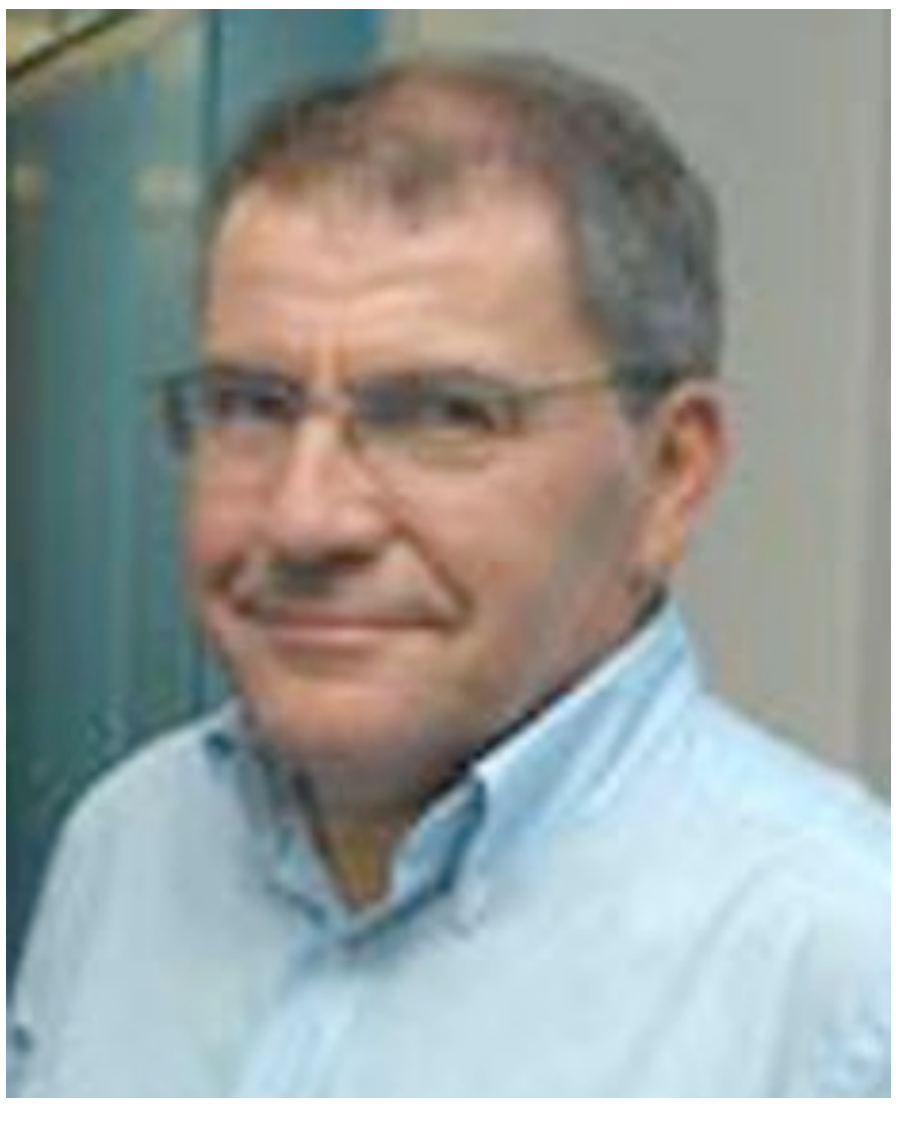

**Jean-François Dartigues** Dr of Medicine, Dr of Sciences, is professor at the University Bordeaux Segalen, France, director of Research group Epidemiology and Neuropsychology of Aging at the INSERM center "Epidemiology and Bio-Statistics", U897. He is Director of Center of Memory Resources and Research of Bordeaux, France. His research is essentially focused on studies and tracking of cohorts of elderly persons. He has been an originator and a principal investigator of Paquid cohort since 1988 and 3C cohort since 1999. He is a member of the board of Alzheimer Plan in France. Since 1988 he has been a member of Project Management group of EURODEM. For his outstanding contribution in Public Health he received Academic Palms decoration in 2006. He is an authors and co-authors of more than 400 research publications.